\documentclass[12pt]{article}
\usepackage{amsmath}
\usepackage[margin=2.5cm]{geometry}

\usepackage{graphicx}
\usepackage{subcaption}
\usepackage{csquotes}

\usepackage[affil-it]{authblk}

\usepackage[normalem]{ulem}

\newcommand{\erf}{\mathrm{erf}\,}
\newcommand{\erfc}{\mathrm{erfc}\,}
\newcommand{\hlight}[1]{\textsl{#1}}

\usepackage{xcolor}
\newcommand{\markthis}[1]{#1}
\usepackage{epsfig}
\usepackage{amsfonts}
\usepackage{fancyhdr}
\usepackage{titlesec}
\usepackage{cite}
\usepackage{hyperref}
\usepackage{minitoc}
\usepackage[nolists, nomarkers]{endfloat}
\usepackage{setspace}
\titleformat{\section}
   {\normalfont\large\bfseries\sffamily}{\thesection}{1em}{}
\titleformat{\subsection}
   {\normalfont\normalsize\bfseries\sffamily}{\thesubsection}{1em}{}

\usepackage{caption}

\newcommand{\runningtitle}{\small Shear bands in materials processing} 
\newcommand{\authorname}{\small K. Viswanathan \emph{et al.}} 
\title{\large \sffamily \textbf{Shear bands in materials processing: Understanding the mechanics of flow localization from Zener's time to the present}}
\author[1]{\normalsize Koushik Viswanathan\thanks{koushik@iisc.ac.in}}
\author[2]{Shwetabh Yadav}
\author[2]{Dinakar Sagapuram\thanks{dinakar@tamu.edu}}
\affil[1]{\small Indian Institute of Science, Bengaluru, India}
\affil[2]{\small Texas A\& M University, College Station, TX}
\date{\small October 05, 2020}
\begin{document}
\maketitle
\thispagestyle{plain}
\noindent\hrulefill
\begin{abstract}
\noindent Shear banding is a material instability in large strain plastic deformation of solids, where otherwise homogeneous flow becomes localized in narrow micrometer-scale bands. Shear bands have broad implications for materials processing and failure under dynamic loading in a wide variety of material systems ranging from metals to rocks. This year marks 75 years since the publication of Zener and Hollomon's pioneering work on shear bands (C. Zener and J. H. Hollomon, \emph{J Appl. Phys.}, 15:22-32, 1944) which is widely credited with drawing the attention of the mechanics community to shear bands and related localization phenomena. Since this landmark publication, there has been significant experimental and theoretical investigation into the onset of shear banding. Yet, given the extremely small length and time scales associated with band development, several challenges persist in studying the evolution of single bands, post-initiation. For instance, spatiotemporal development of strain fields in the vicinity of a band, crucial to understanding the transition from localized flow to fracture, has remained largely unexplored. Recent full-field displacement measurements, coupled with   numerical modeling, have only begun to ameliorate this problem. This article summarizes our present understanding of plastic flow dynamics around single shear bands and the subsequent transition to fracture, with special emphasis on the post-instability stage. \markthis{These topics are covered specifically from a materials processing perspective. } We  begin with a semi-historical look at some of Zener's early ideas on shear bands and discuss recent advances in experimental methods for mapping localized flow during band formation, including direct \emph{in situ} imaging as well as \emph{ex situ}/post-mortem analyses. Classical theories and analytical frameworks are revisited in the light of recently published experimental data. We show that shear bands exhibit a wealth of complex flow characteristics that bear striking resemblance to viscous fluid flows and related boundary layer phenomena. Finally, new material systems and strategies for reproducing shear band formation at low speeds are discussed. It is hoped that these will help further our understanding of shear  band dynamics, the subsequent transition to fracture, and lead to practical `control' strategies for suppressing shear band-driven  failures in processing applications.
\end{abstract}

\noindent\hrulefill
{\sffamily \tableofcontents}
\noindent\hrulefill
\section{Introduction}

Shear banding is a phenomenon that occurs during plastic deformation of several material systems, ranging from metals to rocks, typically under conditions of high strain rate \cite{BaiDodd_1992}. It entails spatial localization of plastic flow in thin zones (shear bands)  even when the remote loading is uniform or homogeneous. Material within the bands can often deform to very large strains ($\sim 5$ or more) with the surrounding regions remaining largely unstrained. The formation and development of shear bands has been of long-standing interest since localized flow zones often act as precursors to catastrophic fracture. This understandably has adverse consequences, particularly in applications such as materials processing (e.g., machining, forming) that involve large strain plastic deformation \cite{shaw1954,backofen1972}. Shear banding is now also recognized as an important mechanism for material failure in high-rate dynamic loading applications such as shock, impact and projectile penetration \cite{clifton2000}.

Shear banding has been known for at least 150 years, dating back to the work of Henri Tresca. In the ensuing period, several excellent reviews and monographs have already been written on the subject, notably focusing on intrinsic material/microstructural factors that lead to band initiation, band properties in various material systems from ductile metals to metallic glasses, and related mathematical modeling \cite{BaiDodd_1992, AntolovichArmstrong_PGMS_2014,Rogers_AnnRevMatSci_1979, Wright_2002,meyers1994}. A historical review of shear bands, particularly focusing on their discovery in the late 19$^{th}$ and early 20$^{th}$ can be found in the articles by Dodd, Walley and co-workers \cite{walley2007, DoddETAL_MetTrans_2015}.

The modern era of shear band research can be dated back to the work of Zener and Hollomon in the 1940s \cite{ZenerHollomon_JApplPhys_1944}, which correctly identified band formation as arising from a plastic instability, analogous to phenomena such as tensile necking, plastic buckling and kinking \cite{Bridgman_1952,nadai_1950, Shanley_1947}. The present review on shear bands is presented to commemorate the 75$^{th}$ anniversary of this pioneering work, and has a modest but important focus. \markthis{We discuss the mechanics of shear banding in the oft-neglected context of materials and deformation processing, for this was also the setting of Zener and Hollomon's original work. We show that apart from their practical importance, deformation processes like cutting also provide the ideal framework to study the two-stage process of band formation involving distinct initiation (\emph{nucleation})  and  development (\emph{growth}) phases. In particular, we discuss recent developments in \emph{in situ} imaging and full-field deformation measurements, coupled with unified analytical models. Our work is restricted to polycrystalline metals. The vast literature on metallic glasses is beyond the scope of the present article.}


We begin with a semi-historical revisit of Zener and Hollomon's contributions and explore their original ideas on shear bands (Sec.~\ref{sec:ZenerHistory}). The primary experimental observations pertaining to banding are then briefly presented---these are features that any theory should satisfactorily explain (Sec.~\ref{sec:shearBandFeatures}). Following this, we discuss present consensus on why shear bands form (Sec.~\ref{sec:shearBandInitiation}), covering both continuum-scale instability analyses as well as possible microstructure-level mechanisms. The  post-initiation band development and kinematics of localized plastic flow are discussed next (Sec.~\ref{sec:shearBandDevelopment}). Transition from localization to fracture is reviewed in Sec.~\ref{sec:fractureTransition}, delineating some possible mechanisms for crack initiation and growth. Practical consequences for materials processing and failure are discussed in Sec.~\ref{sec:applications}. We conclude with some open questions and future research directions in Sec.~\ref{sec:discussion}.


\section{Zener's contributions and the 1944 paper}
\label{sec:ZenerHistory}

\begin{figure}
  \centering
\begin{subfigure}[b]{0.75\textwidth}
\centering
  \includegraphics[width=0.9\textwidth]{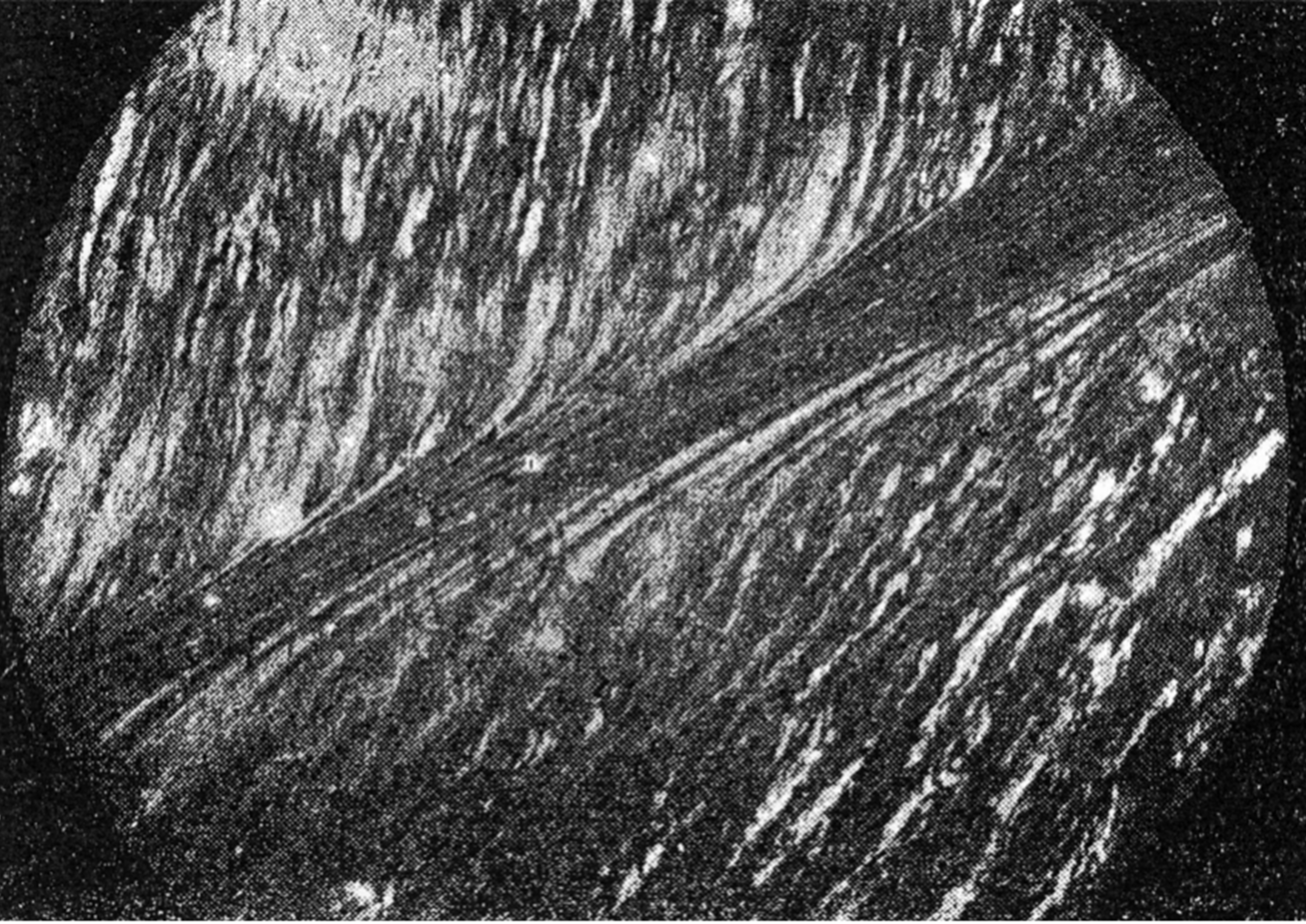}
  \caption{}
  \end{subfigure}
  
  \begin{subfigure}[b]{0.75\textwidth}
  \centering
  \includegraphics[width=0.9\textwidth]{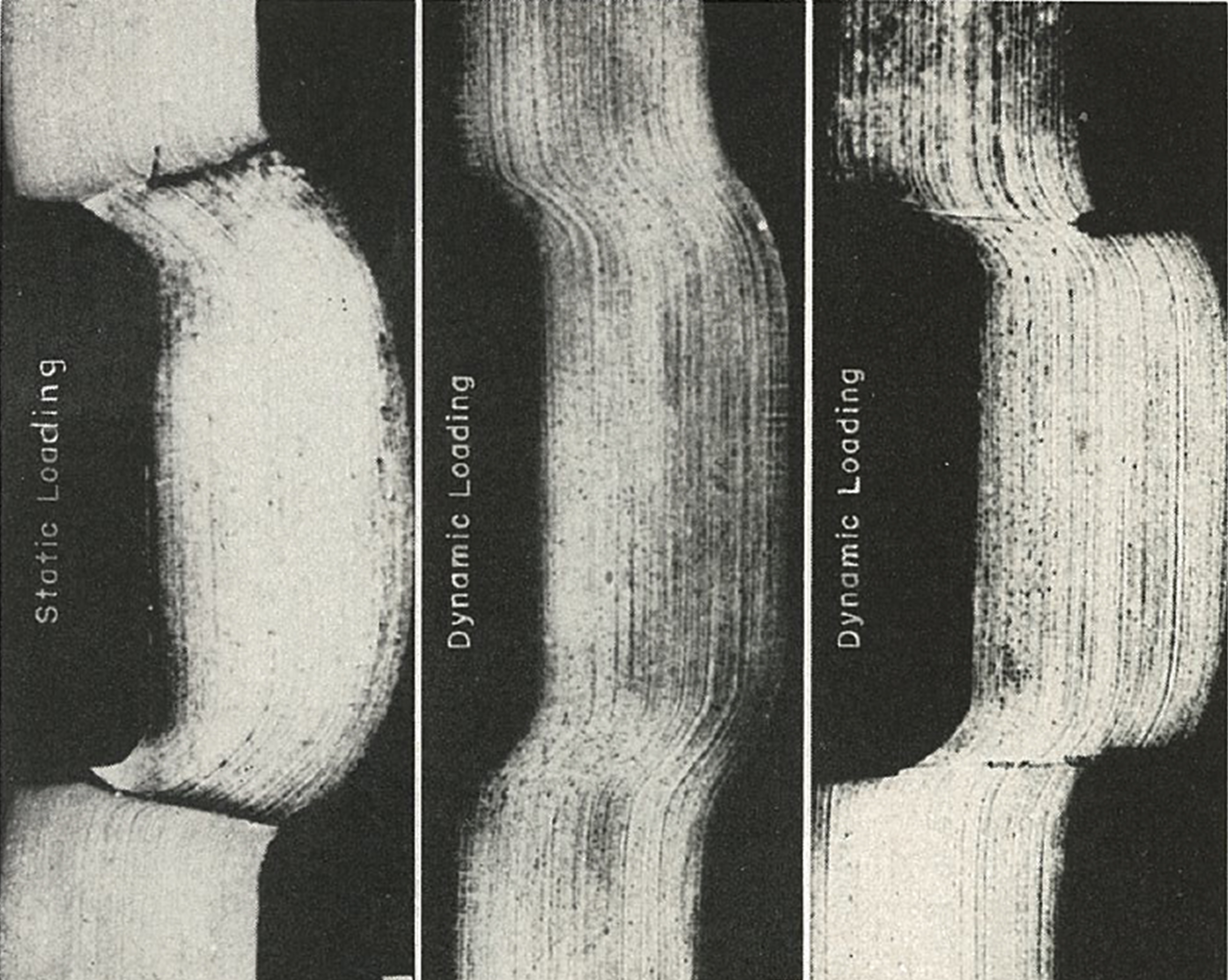}
  \caption{}
  \end{subfigure}
  
  \caption{(a) Optical micrograph showing structure of a shear band formed in austenitic steel after impact by a 50 kg weight from a height of 2.5 m \cite{dodd2014}. (b) Transition in the plastic flow from homogeneous to localized type with increasing rate of deformation in punching of armor steel plate \cite{zener1948}. In static loading, deformation is spread out, while in the dynamic loading case, deformation is localized in the immediate vicinity of the punching surface. Partial incomplete shear along the localized shear surface in these experiments was achieved by adjusting the height from which the punch was dropped.  }
  \label{fig:historical}
\end{figure}

The fact that plastic flow in metals can, under certain conditions of rapid straining, become spatially concentrated in thin zones, has been known since well before Zener's time. The earliest recorded observations of these zones were perhaps made by Henri Tresca as part of an extensive investigation of plastic flow in metals. Tresca presented his observations to the Institution of Mechanical Engineers in Paris in 1878 in a lengthy paper comprising some 28 pages and 65 figures \cite{tresca1878}. In particular, Tresca's observations of hammer forging in iridized platinum showed sharp X-shaped localized zones of high temperature. These were argued to arise from preferential plastic shearing, which increased the local temperature to \lq red hot\rq . In virtually identical experiments with steels nearly 40 years later, the same phenomenon was rediscovered by Harold Massey at Manchester, and the crosses were termed \lq heat lines\rq \cite{massey1921}. A more dramatic manifestation of shear localization was parallely discovered by a Russian metallurgist Kravz-Tarnavskii \cite{DoddETAL_MetTrans_2015} in a little-known work on dynamic compression by a drop hammer. The resulting \lq peculiar white bands\rq\ were the result of local heating, leading to an austenite to martensite transformation, and followed by rapid quenching. Similar observations of white bands  during rapid upsetting of steels were later presented by Davidenkov and Mirolubov in 1935 \cite{dodd2014} and Trent in 1941 \cite{trent1941}. Micrographs published by Kravz-Tarnavskii, Davidenkov and Mirolubov, reproduced in Fig.~\ref{fig:historical}(a), perhaps represent the first photographic evidence of shear bands.  It is interesting to note that the above noted observations of shear bands, including that of Adcock \cite{adcock1922}, were all made in the context of metals processing.

This, then, was the general state of affairs in the 1940s when Zener and Hollomon first presented a mechanics-based analysis. The setting for this work was Watertown Arsenal near Boston, a US Army laboratory which played a key role in the development of materials science and technology during World War II. According to Zener, the arsenal at that time was concerned with the design and processing of armor plate to resist penetration, and of projectiles to penetrate armor \cite{gilman1994}. Zener was brought in as the head of the \emph{metallurgical physics} department, while Hollomon (who had just received his PhD from MIT) was made the head of the \emph{physical metallurgy} department. This collaboration between the two---with the agreement that Hollomon would teach Zener metallurgy, while the latter would teach him physics---was certainly unusual for that time.

Their landmark paper, titled \lq Effect of strain rate upon plastic flow of steel\rq\ \cite{ZenerHollomon_JApplPhys_1944} is remarkable for its insight on the underlying mechanics. For the first time, shear localization was cast in the form of a plastic instability problem, resulting from a competition between material strain hardening and softening due to local deformation-induced temperature rise. This paper continues to attract some 150+ citations per year, and is widely credited with also drawing attention to shear banding phenomena in ordnance/ballistic impact applications in addition to materials processing.

Interestingly, the results on shear localization appear only at the end of this 1944 paper, in a section titled \lq Dynamical consequences of adiabatic relations.\rq\ The most important idea from this work is that at sufficiently high strain rates, when the heat generated due to plastic deformation has insufficient time for conduction, the deforming material becomes hot enough that its flow stress can decrease with strain; this is in contrast to the usual strain hardening at quasi-static strain rates. The result of this effective ‘strain softening’ is an autocatalytic effect,  wherein  once preferential slip begins along any plane, further deformation continues along the same plane. In fact, an almost identical instability mechanism/criterion for shear bands, and its analogy to the case of necking in a tensile specimen, was proposed by Davidenkov and Mirolubov in 1935 \cite{dodd2014}. However, the Zener-Hollomon work was almost certainly independent of this result, since it is only recently that this early Russian work has become known \cite{DoddETAL_MetTrans_2015}.

Zener and Hollomon dramatically demonstrated shear localization using a die and punch set-up, where the punch was made to penetrate about half-way into a steel plate at 3 m/s (strain rate $\sim 2 \times 10^3$ /s) by dropping a weight. Just as they had predicted, this produced a thin and well-defined band of about 30 $\mu$m in thickness and a shear strain of nearly 100 right at the surface where the plug had detached from the plate; see Fig.~\ref{fig:historical}(b). Under these adiabatic conditions, calculations predicted a temperature rise of $\sim 1000$ $^\circ$C  for a shear strain of  5, so the band would have been subject to substantial temperature rise, consistent with previous reports. 

Besides the results on shear bands, another well-known idea to come out of this collaboration was that the stress-strain relation in metals depended upon strain rate ($\dot\varepsilon$) and absolute temperature ($T$) only through a single parameter $Z = \dot\varepsilon \exp(Q/RT)$, where $R$ is the universal gas constant and $Q$ the activation energy \cite{ZenerHollomon_JApplPhys_1944,zener1944a}. This result appears to be part of an ongoing search to determine a universal equation of state for plastic deformation of metals, analogous to that for ideal gases \cite{gray1995}. The implication of this result is that the effects of increasing strain rate on the deformation can be studied under isothermal conditions via equivalent reduction in the temperature. While Zener and Hollomon had proposed this concept in an earlier paper \cite{zener1944a}, the experimental tests and associated data reported in the 1944 paper were particularly important in demonstrating the equivalent effects of $\dot\varepsilon$ and $T$. The use of the parameter $Z$, referred to as the Zener-Hollomon parameter, is now become pervasive in various domains of materials science, most importantly, creep \cite{weertman1986}.

It is noteworthy that Zener and Hollomon \emph{a priori} predicted and demonstrated catastrophic shear localization purely based on adiabatic arguments, in contrast to other works until that point (by Tresca, Massey and others) where the phenomenon was discovered largely by accident. The 1944 paper thus represents a good example of how physical insight, when coupled with clever experiments, can lead to major advances. An excerpt from Zener's interview that sheds some light on the development of these ideas is provided in Appendix \ref{sec:appendix}. 

In addition to providing a physical basis for catastrophic slip and failure and triggering large-scale interest in shear banding phenomena in diverse material systems, the 1944 paper and a succeeding article by Zener in 1948 \cite{zener1948} are also noteworthy for their ability to correlate a number of apparently disparate phenomena. For example, the idea that catastrophic slip is a plastic instability  accompanied by a large local temperature rise has played an important role in elucidating discontinuous yielding and load drops at very low-temperature deformation of metals \cite{basinski1957,chin1964a}; formation of \lq cup-cone\rq\ shaped fractured surface in tensile loading of ductile metals \cite{chin1964,backofen1972}; serrated (shear localized) chip formation in high-speed machining \cite{shaw1954}; microscopic processes leading to detonation of explosives during dynamic loading \cite{winter1975}; the mechanism of material removal in erosion \cite{ChristmanShewmon_1979, ShewmonSundararajan_1983}; and even seismic faulting at a much grander scale \cite{poirier1980,rice2006}.

\section{General features of shear bands}
\label{sec:shearBandFeatures}

It is best to begin by recalling the primary phenomenological features of shear bands that any valid theory must aspire to explain. We shall limit ourselves to ductile polycrystalline metals which usually deform homogeneously under quasi-static conditions but exhibit localized flow at high strain rates. It should be remembered that this phenomenon is not limited to any particular loading configuration or alloy system, but has been observed in a range of  engineering problems including machining \cite{recht1964, komanduri2002,sagapuram2020}, forging/upsetting \cite{tresca1878, johnson1964}, blanking/piercing \cite{dodd1983}, torsion \cite{johnson1983,marchand1988,giovanola1988 } and impact \cite{wingrove1973,timothy1985}.

Some distinguishing characteristics of shear bands formed under these conditions of high strain rate are:
\begin{enumerate}
\item \hlight{Occurrence in different material systems (pure metals, substitutional/multi-phase alloys) and crystal structures (\emph{fcc}, \emph{bcc} and \emph{hcp})}. While most pronounced in alloys characterized by high strength and low thermal diffusivity (e.g., steels, titanium and nickel-based superalloys),  banding has been also reported in ductile pure metals (e.g., Cu), albeit at very large strains \cite{hatherly1984}.
  
\item \hlight{Very small thickness}, usually in the range of $5$-$50\,\mu$m \cite{BaiDodd_1992,timothy1985}, with this dimension being much smaller than specimen dimensions or any other length scales.

\item  \hlight{Relative insensitivity to local stress state}: Band widths are largely independent of local stress state but show strong dependence on material properties \cite{BaiDodd_1992}. High-strength alloys such as titanium and steels exhibit narrow and well-defined bands, while more diffuse bands are typical of very ductile \emph{fcc} metals such as aluminum \cite{timothy1987}.
  
  \item \hlight{Extreme deformation  within the  band}: For instance, shear strains in the range of $5$-$100$ have been reported based on marker/grid measurements or post-mortem observations of deformed microstructural features \cite{manion1969,sagapuram2018a,ZenerHollomon_JApplPhys_1944}. 
  
\item \hlight{Very small time scales ($< 100$ $\mu$s)} are associated with shear band development, as deduced using high-speed imaging experiments \cite{zhou1996,roessig1998,sagapuram2016}.  
  
  \item \hlight{Local strain rates much larger than nominally applied values}: While direct measurements of shear band strain rates are  scarce, local estimated strain rates ($10^4$-$10^7$ /s) are 1-2 orders of magnitude higher than the nominal applied strain rates. Similarly, the local temperature rise can be substantial, of the order of several hundred degrees, depending on the material and local strain rate \cite{zhou1996, hartley1987}. The extreme case of melting within a band has been also postulated in some cases \cite{hartmann1981,thompson1970}.
    
  \item \hlight{Unrestrained propagation independent of microstructural obstacles}: Band formation is seldom deterred by microstructural features such as grain or phase boundaries \cite{Wright_2002}. This must be compared with other instabilities, such as plastic buckling, whose evolution is often dictated by the local grain structure \cite{YeungETAL_ProcNatAcadSci_2015, UdupaETAL_ProcRoySocA_2017}.  

    \item \hlight{Non-crystallographic nature}: Orientation of shear bands is often determined by the macroscopic stress state and not by local crystallographic directions. This  distinguishes them from microscopic localization phenomena such as slip bands, kink bands and deformation twins.




\item \hlight{Highly refined microstructure}, composed of very fine cells/grains ($< 0.1$ $\mu$m). While this was recognized very early on by Kravz-Tarnavskii (based on the ‘non-etchable’ nature of shear bands), it was only with the later developments in TEM imaging that the microstructural features within a band could be fully resolved \cite{glenn1971,wingrove1971, meyers1986,sagapuram2016,landau2016}.

\end{enumerate}

\begin{figure}
  \centering

  \includegraphics[width=1\textwidth]{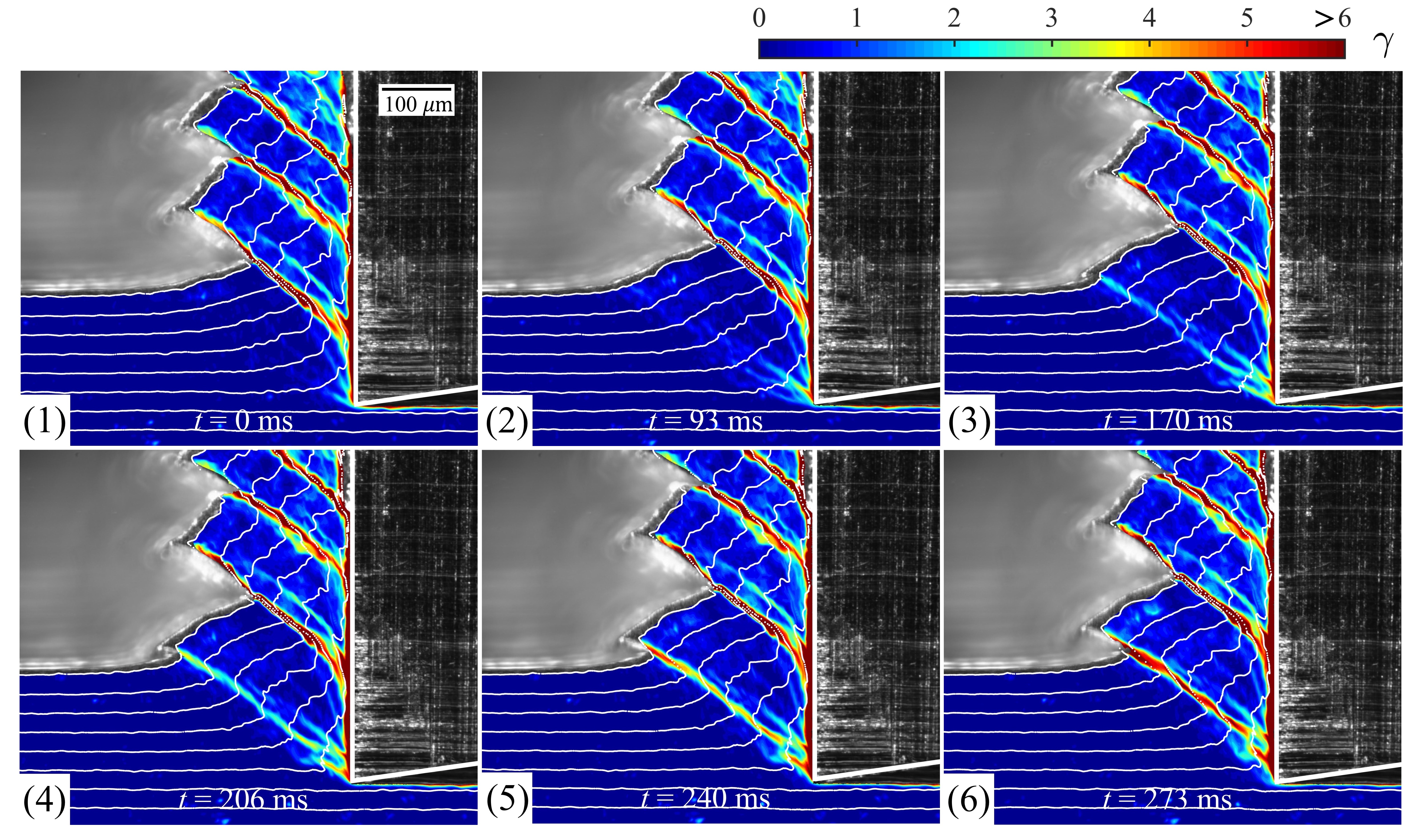}
  
  \caption{High-speed images showing shear band initiation, propagation and associated strain field development in two-dimensional cutting of a low melting point bismuth-based alloy (Wood's metal). The streaklines and full-field shear strain information  was obtained using PIV image processing technique. The shear band development occurs by band nucleation at the tool tip (frame 1), propagation of the band front towards the sample free surface (frames 2-3), followed by macroscopic sliding along the band plane (frames 4-6).  This last stage of  sliding accounts for most of the strain localization around the band.} 
  \label{fig:sb_initiation}
\end{figure}

In addition to these general characteristics, recent developments using high-speed \emph{in situ} imaging have shown that the dynamics of shear band formation can be clearly demarcated into two distinct phases, see Fig.~\ref{fig:sb_initiation}. This figure presents a sequence of frames of two-dimensional cutting of a low melting point fusible alloy, taken using high-speed photography and processed using digital image correlation techniques. Overlaid on these frames are computed plastic strain fields that show the extent of the localization. Band formation clearly occurs in two phases:
\begin{enumerate}
\item Nucleation of a single band as a local weak interface starting at the  tool tip (frames 1-3). This first phase establishes the band orientation and path. 
  \item Localized \lq sliding\rq\ along the nucleated band (frames 4-6). It is clear from the figure that substantial plastic flow occurs across the band only during this second phase.
  \end{enumerate}
  
  Recent work has found this two-phase mechanism of shear band development to be largely independent of both the deformation geometry as well as microstructural features, having been observed in a range of other alloy systems \cite{sagapuram2016,sagapuram2018b,yadav2020}.

In the following sections, we discuss the mechanics of shear banding from this \lq two-phase\rq\ viewpoint, where aspects pertaining to shear band initiation, i.e., onset of instability, are decoupled from the strain-intensive and largely material independent growth phase. Issues related to fracture along the band, which usually occurs late into the growth phase, are reviewed subsequently.


\section{Why do shear bands form?}
\label{sec:shearBandInitiation}

The first step in the formation of a fully developed shear band is the initiation phase, where a weak interface---the incipient band---is nucleated under homogeneous loading. Most thermo-mechanical investigations of shear banding have focused exclusively on predicting the onset of this initiation. This is understandable, since strain localization is all too often a precursor to inevitable fracture so that initiation predictions are of practical importance, and have spurred both modeling and experimental efforts. In the following, we will focus primarily on analytical models and will leave out the vast literature on finite element simulations of band formation. \markthis{A good starting point for beginning a study of the latter is the article by Needleman \cite{needleman1992a}.}

\subsection{Continuum-scale analyses based on phenomenological considerations}

Given that shear bands occur commonly and with seemingly complete periodicity in many processing applications, it is not surprising that their initiation can be cast as a classical thermomechanical instability problem. In this context, it is instructional to first consider the possibility of predicting their initiation using only continuum-scale, mostly isotropic, criteria. Following the original work of Zener and Hollomon, Recht formulated an adiabatic criterion in the context of metal machining \cite{recht1964}. This is quite analogous to how one describes the necking instability in tension, where deformation localizes when \lq geometric\rq\ softening (due to area reduction) is balanced by strain hardening. As noted earlier, the corresponding balance for shear band initiation is postulated to occur between thermal softening and strain hardening, as
\begin{equation}
  \label{eqn:onsetCriterion}
0 \leq \frac{\frac{d\tau}{d\varepsilon}}{-\frac{d\tau}{d\theta}\frac{d \theta}{d\varepsilon}} \leq 1
\end{equation}
where $\tau, \varepsilon, \theta$ are the material's shear strength, applied strain and temperature. Initially, the strain hardening term $\frac{d\tau}{d\epsilon}$ term dominates the deformation response. If the thermal conductivity is low enough, local heat generation can cause the softening term $\frac{d\tau}{d\theta}$ to eventually catch up. This relation therefore quantifies how quickly a material must harden or soften for band initiation to occur. Analogous analyses incorporating strain rate and crystallographic texture effects were later presented by Backofen,  Argon and Dillamore \cite{backofen1972, argon1973, dillamore1979}. These criteria have often been criticized as being quantitatively deficient, especially in situations where banding is known to be dependent on the material's yield behavior. For instance, the balance in Eq.~\ref{eqn:onsetCriterion} does not have anything to say on the nature of the yield surface, a problem that is particularly acute when the loading is multi-axial. 

Assuming rate-independent behavior, Rice \cite{rice1976} proposed a general continuum-scale framework for band initiation as a function of the material's constitutive behavior. For materials with piecewise linear constitutive laws, it was shown that initiation was the result of a local bifurcation problem . In addition, Rice also described an alternative dynamic localization problem, similar to Hill's treatment of acceleration waves in solids \cite{Hill_1962}. Here a local discontinuity in material acceleration first propagates in a plastically deforming material. When the propagation velocity of such a discontinuity front becomes zero, for some suitable combination of material and deformation parameters, the flow localizes in the form of stationary shear bands. Corresponding calculations for rigid-plastic and elastic-plastic materials were presented in Ref.~\cite{rice1976}, and analogous to Recht's result (Eq.~\ref{eqn:onsetCriterion}), Rice obtained a condition on the hardening rate for the onset of banding.

As a prototypical bifurcation problem, shear band initiation has also attracted the attention of physicists and applied mathematicians working on nonlinear dynamics. Here, the requisite nonlinearity comes from the material's thermophysical constitutive law, involving exponential terms \cite{johnson1983a}. In a simple shear configuration (cutting), it can be shown that periodic banding arises due to a Hopf bifurcation from the steady homogeneous deformation state \cite{BurnsDavies_1997}. The cutting speed or strain rate plays the role of a bifurcation parameter, consistent with earlier discussions of rate effects \cite{recht1964, rice1976}. The bifurcation results in the formation of a limit cycle, whose oscillation period directly provides the band spacing. It has also been postulated that this type of shear banding is the basis for the \lq segmented chip\rq\ often seen in high-speed machining \cite{ViswanathanETAL_CIRPAnnals_2017}.

\markthis{Nonlinear dynamics analyses of this kind have to necessarily make restrictive approximations about spatial variations. This can be overcome by performing a spatio-temporal analysis, as proposed by Clifton, Molinari and co-workers \cite{clifton1980,clifton1984,molinari1987}, who presented a perturbation analysis of the stability of a solid under simple shear, including the effects of heat conduction, strain hardening, rate sensitivity and thermal softening. This method has been used with great success to predict whether or not a shear banding instability will occur for a given constitutive law, and  the critical strain at which the localization turns catastrophic,  as a function of the initial perturbations (e.g., in the form of geometrical defects or nonuniform initial temperature and strain fields) as well as the intrinsic material parameters. Since perturbation analysis also provides information on  the dominant wavelengths corresponding to the instability mode, further predictions could be made for band spacing from the fastest growing wavelengths in materials both with and without strain hardening \cite{Molinari_1997}. The results from these analyses were found to be in reasonable agreement with experiments. 
An analogous 1D viscoplastic analysis of shear band thickness, spacing between multiple bands, and thickness of an equivalent thermal boundary layer was presented by Wright \cite{Wright_2002}. In the latter context of thermal layer, mention should be also made of Shawki and co-workers \cite{shawki1992} who introduced an energy criterion for the onset of shear localization in thermal viscoplastic materials (no strain dependence) and presented a unified analytical description for the full localization history from a kinetic energy viewpoint. This framework was used to argue that the thermal length scale introduced by heat diffusion can be correlated with the finite shear band thicknesses observed in experiments \cite{cherukuri1995}.}



In all of these continuum-level models, microstructural contributions, such as grain- or dislocation-scale structures, to the onset of shear banding were either entirely ignored or considered by including a suitable term in the constitutive relations. \markthis{In this context, it is pertinent to mention that often thermal-induced softening alone is not sufficient to balance strain hardening and recourse is often made to a continuum-scale damage parameter, depending on the local stress-state. These are unfortunately difficult to compute or evaluate analytically, and require the use of finite element simulations. The interested reader can find some pointers in Ref.~\cite{Giglio_2012}}. 

\subsection{Microstructural factors influencing band initiation}


While Zener's work and the subsequent continuum-scale analyses described above provide a macroscale rationale for why a material instability should occur, they do not shed any light on how band initiation is physically realized on the microscale. In the same spirit as Zener, Backofen \cite{backofen1972} states:
\begin{displayquote}
  The condition for initiating unstable flow after any amount of prior straining is that somewhere in the material, the next increment of strain-induced hardening be cancelled out by an accompanying strain-induced softening, then further straining will tend to concentrate in the plane where resistance to continuous flow is first lost.
\end{displayquote}
In the ensuing paragraphs, we shall concern ourselves with a discussion of how such a plane---the nucleated \lq weak interface\rq\ of the continuum-scale analyses---may be formed from a physical standpoint.


First, it is clear that band initiation must be accompanied by a marked reduction in the local flow stress with increasing temperature, for this will provide the driving force for subsequent deformation localization. Several microscopic mechanisms can contribute to this flow stress reduction: stress-induced climb of dislocation networks \cite{ramalingam1973}, dynamic recrystallization \cite{rittel2008,landau2016},  phase transformations and even melting \cite{lewandowski2006}. All these mechanisms are inherently diffusion-driven, so their rates are controlled by the corresponding diffusion coefficients: lattice diffusion coefficient (climb-controlled dislocation motion, phase transformations) \cite{frost1982, christian_2002} and grain boundary diffusivity (dynamic recrystallization) \cite{gottstein2009}. Therefore, any microscopic explanation that invokes these mechanisms should also be able to satisfactorily account for the very small time scales associated with shear band initiation. However, detailed studies of the kinetics of these thermally-activated processes, and corresponding comparisons with intrinsic shear band nucleation time scales are very few, save for the case of dynamic recrystallization \cite{hines1997}. However, the question of whether recrystallization is universally observed in every instance of shear banding as yet remains unanswered.

This brings us to alternative athermal mechanisms for shear band nucleation. For example, one can envisage the formation of a local weak interface as arising from a purely mechanical instability involving sudden \lq burst\rq\ of dislocations piled up at moderately strong barriers. To illustrate, consider the process of dislocation jumps across potential barriers, such as second-phase particles and forest dislocations, with $\tau_c$ denoting the average necessary shear stress for this process. If the average applied stress is less than $\tau_c$, dislocation can still occur via thermal fluctuations, provided sufficient time is made available. However, at high strain rates, the external loading rate can be significantly ahead of the material's internal plastic strain rate so that when the stress locally reaches the critical value $\tau_c$, piled-up  dislocations are released along a plane as a sudden burst. Of course, if the temperature rise due to this  pile-up breakthrough is sufficiently high to overcome the strain/strain-rate hardening, the localization results along the plane of released pile-ups according to Zener's criterion. Shear band onset mechanisms based on this type of dislocation \lq avalanche\rq\ have, in fact, been proposed by Gilman \cite{gilman1994}, Lee and Duggan \cite{lee1994}, and Armstrong and co-workers \cite{AntolovichArmstrong_PGMS_2014,antolovich2014}, albeit in slightly different forms. Experimental evidence supporting this idea and demonstrating band nucleation at a critical stress has also been presented recently \cite{yadav2020a}. The above dislocation picture can help us understand several important observations from a physical viewpoint, such as the prevalence of shear banding in a range of crystalline metals and alloys; short time scales for band initiation; and increased propensity at high strain rates, low (cryogenic) temperatures and very high levels of cold-work \cite{hatherly1984}. 

As a final note on other athermal softening mechanisms, we mention two: crystallographic texture and void formation. The former was first suggested by Dillamore \cite{dillamore1979}, and involves local lattice reorientation during deformation, resulting in a \lq softer\rq\ texture along a weak interface. As a result, further strain localization occurs along this weak plane, and this then is where the band actually forms. In fact, this mechanism appears to be the primary contributor to shear band development in highly-deformed metals under nearly isothermal/quasi-static conditions, such as in cold-rolling. On a somewhat larger scale, the initiation and growth of voids may also act as an additional source for softening \cite{BaiDodd_1992,vyas1999,woodward1979,needleman1992}, although the fact that shear bands in  ductile metals are usually free of voids indicates that this mechanism is perhaps important only in alloys having limited ductility/workability.

\section{Post-initiation development of shear bands}
\label{sec:shearBandDevelopment}

It is clear that once banding is initiated and a weak interface is formed somewhere in the material, further displacements and plastic flow are essentially confined to the immediate vicinity of this plane. We now discuss some of these post-initiation aspects of shear bands, divided broadly into flow kinematics (based on displacement/strain measurements) and thermal effects (based on heat diffusion models).

\subsection{Kinematics of the displacement/strain fields}

\begin{figure}
  \centering
\begin{subfigure}[b]{1\textwidth}
\centering
  \includegraphics[width=1\textwidth]{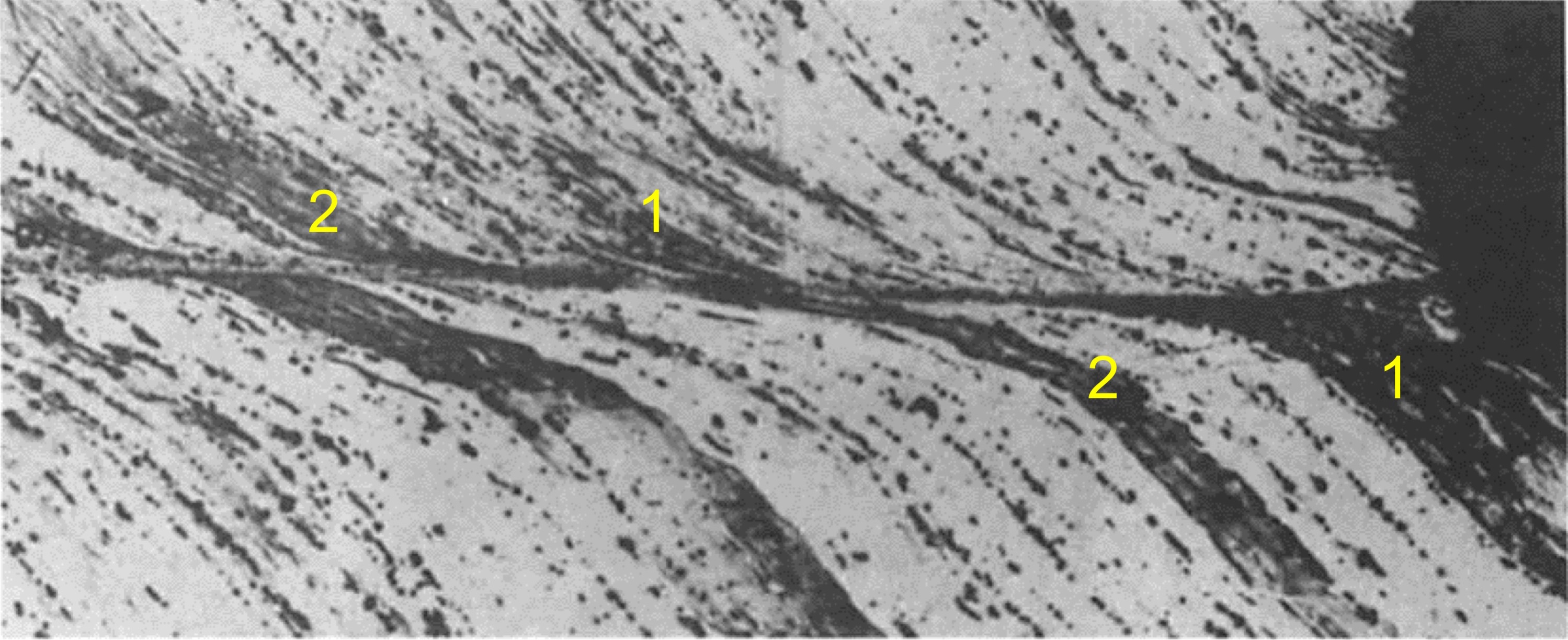}
  \caption{}
  \end{subfigure}
  
  \begin{subfigure}[b]{1\textwidth}
  \centering
  \includegraphics[width=1\textwidth]{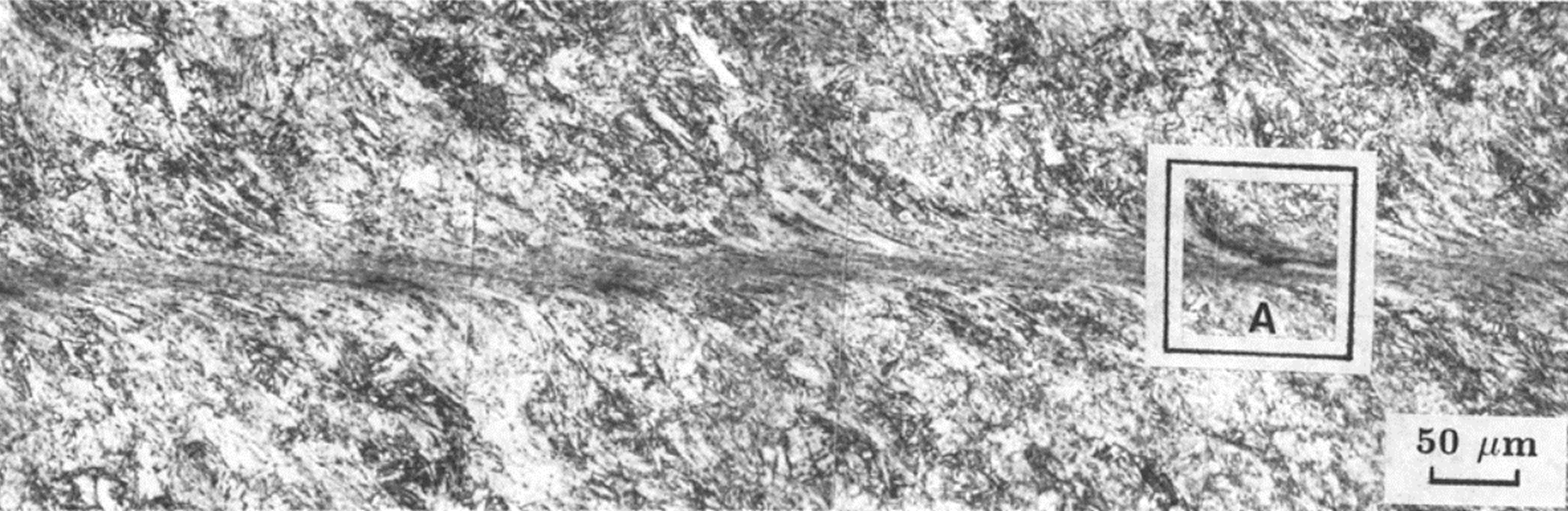}
  \caption{}
  \end{subfigure}
  
  \caption{Optical micrographs showing large localized displacements and boundary layer-type structure in the vicinity of shear bands: (a) shear band formed in hardened (2014 T6) aluminum alloy after projectile impact \cite{wingrove1973}, and (b) shear band in HY-100 steel formed during high-speed torsional test (strain rate $\sim 1000$ /s) \cite{cho1990}. Note the striking similarity in displacement/flow profiles in both the cases despite different material system and loading configuration.
 }
  \label{fig:bandVicinityFlow}
\end{figure}

Localized flow features that are typically seen in the vicinity of a dynamically formed shear band are shown in Fig.~\ref{fig:bandVicinityFlow}. Figure.~\ref{fig:bandVicinityFlow}(a) shows a narrow shear band formed at the corner of an impact crater and extending deep into the specimen in a 2014 aluminum alloy (T6 aged condition). Large localized displacements across the narrow band (measured strain $\sim 100$) are evident from the relative positions of the dark zones marked 1 and 2 (presumably segregated second phase) on either side of the band \cite{wingrove1973}. A similar shear band formed in high-speed torsion of a low alloy carbon steel (HY-100) is  shown in Fig.~\ref{fig:bandVicinityFlow}(b). The plastic flow profiles in both cases are clearly very similar, save for a scaling factor.

Furthermore, the principal features of the shear band flow seen in Fig.~\ref{fig:bandVicinityFlow}---large displacements across the band (without fracture), steep but continuous deformation  gradient perpendicular to the band, and extremely slender shearing zone with respect to its length---are now known to be common characteristics of most, if not all, shear bands \cite{samuels1978, StockThompson_1970, lee1994, timothy1987, Rogers_AnnRevMatSci_1979,ZenerHollomon_JApplPhys_1944}. Since the final displacement profile is ultimately determined by post-initiation kinematics, these observations suggest that localized flow evolution following band initiation may be governed by a common mechanism,  even though the underlying initiation mechanisms  may be material-specific (Sec.~\ref{sec:shearBandInitiation}).

The question of how displacement, velocity and temperature fields evolve around a nucleated band is now quite pertinent. A detailed understanding of these mechanisms, and their dependence on the   material properties and/or external loading factors,  is crucial for understanding the stress collapse and the eventual transition to fracture. Moreover, as pointed out by Duffy \cite{duffy1992}, complete determination of flow history in terms of strain, strain rate, temperature, and band thickness evolution, perhaps represents the most valuable information for validating model predictions. Unfortunately, obtaining this type of \emph{in situ} measurement is severely challenging, given  the extremely small time scales as well as the fact that  the location of shear bands is usually not known \emph{a priori}. Consequently, experimental investigations of spatio-temporal dynamics following band initiation are extremely sparse.

The first attempt at following the deformation process associated with shear banding should be attributed to Wingrove \cite{wingrove1973}. His ingenious \lq quick-stop\rq\ technique demonstrated  the progression of shear band formation during penetration of an aluminum alloy plate by stopping the penetration at various stages, followed by post-mortem metallographic examination. Furthermore, a \lq stepped\rq\ projectile was used to independently control the penetration depth so that different time-snaps of the plugging process could be simulated by \emph{a priori} controlling projectile dimensions. The sequence of steps highlighted in this study---band initiation at the projectile corner (a stress concentration); propagation of the band into the plate thickness and until the back surface; development of a crack at the back surface and its propagation back along the shear band plane to cause final detachment of the plug---pretty much encompass all the key attributes of shear localization-driven failure as we know it today.  We also note that similar quick-stop experiments to achieve incomplete shear along a band by arresting the deformation were made much earlier by Zener himself \cite{zener1948}, although details of where and how  bands formed  were not discussed. Additionally, key aspects of the strain localization, for example displacement/strain profiles and their evolution, were not resolved in Zener's work.

These details were first outlined, in parallel, by Marchand and Duffy \cite{marchand1988} and Giovanola \cite{giovanola1988}, with the help of high-speed \emph{in situ} photography techniques, and they represent the first direct observations of shear band dynamics. In both these studies, the initiation and growth of shear bands under dynamic torsion ($\sim 10^3$ /s strain rates) of thin-walled tubes, were inferred by tracking the deformation of a deposited grid pattern. Marchand and Duffy's \emph{in situ} experiments observed the deformation as a function of position around the circumference of the specimen, revealing band nucleation at a single site, it’s rapid propagation ($\sim 500$ m/s), followed by fracture along its path, identical with earlier results \cite{wingrove1973}.

\begin{figure}
  \centering
\begin{subfigure}[b]{1\textwidth}
\centering
  \includegraphics[width=1\textwidth]{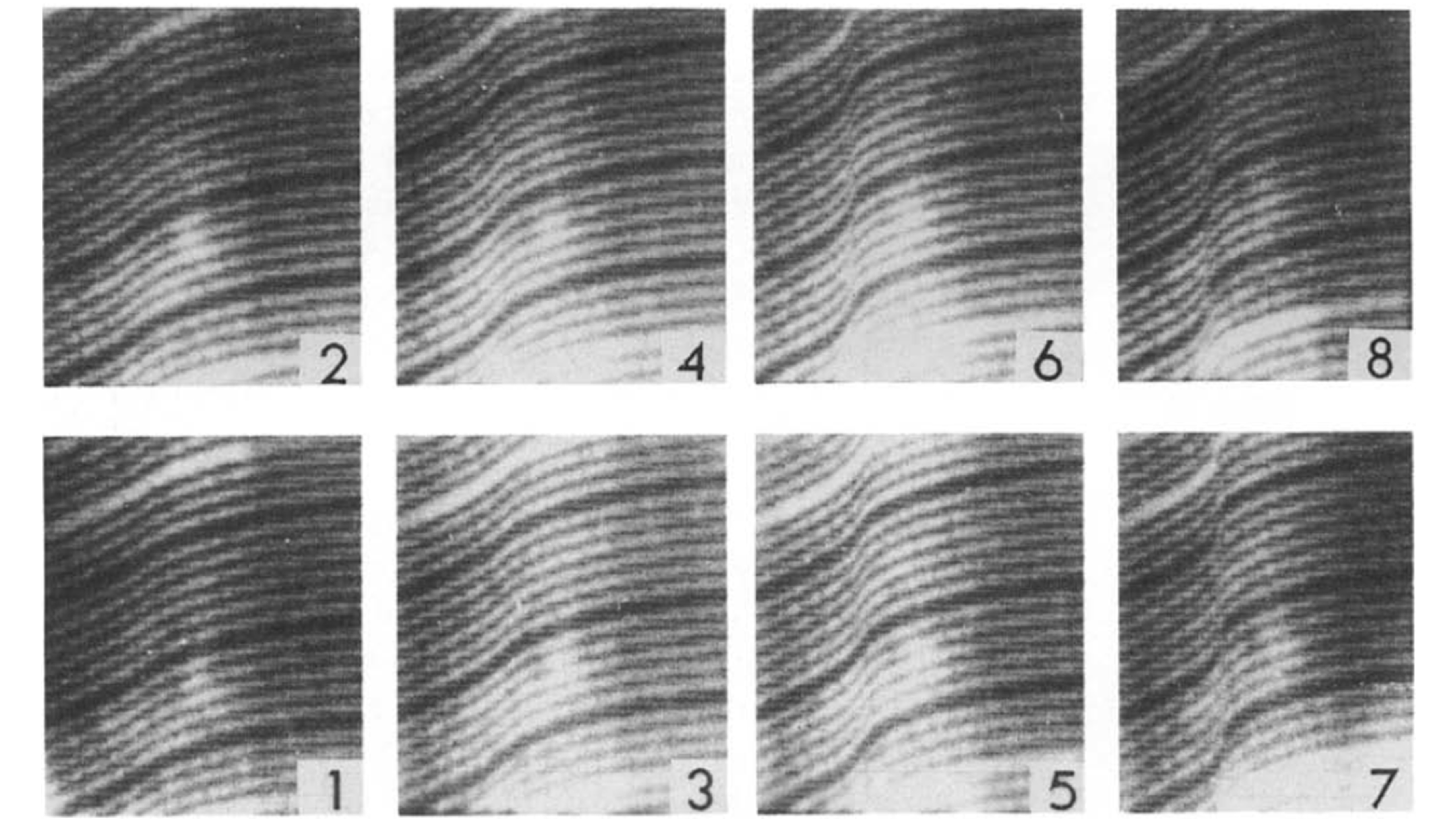}
  \caption{}
  \end{subfigure}
  
  \begin{subfigure}[b]{1\textwidth}
  \centering
  \includegraphics[width=0.8\textwidth]{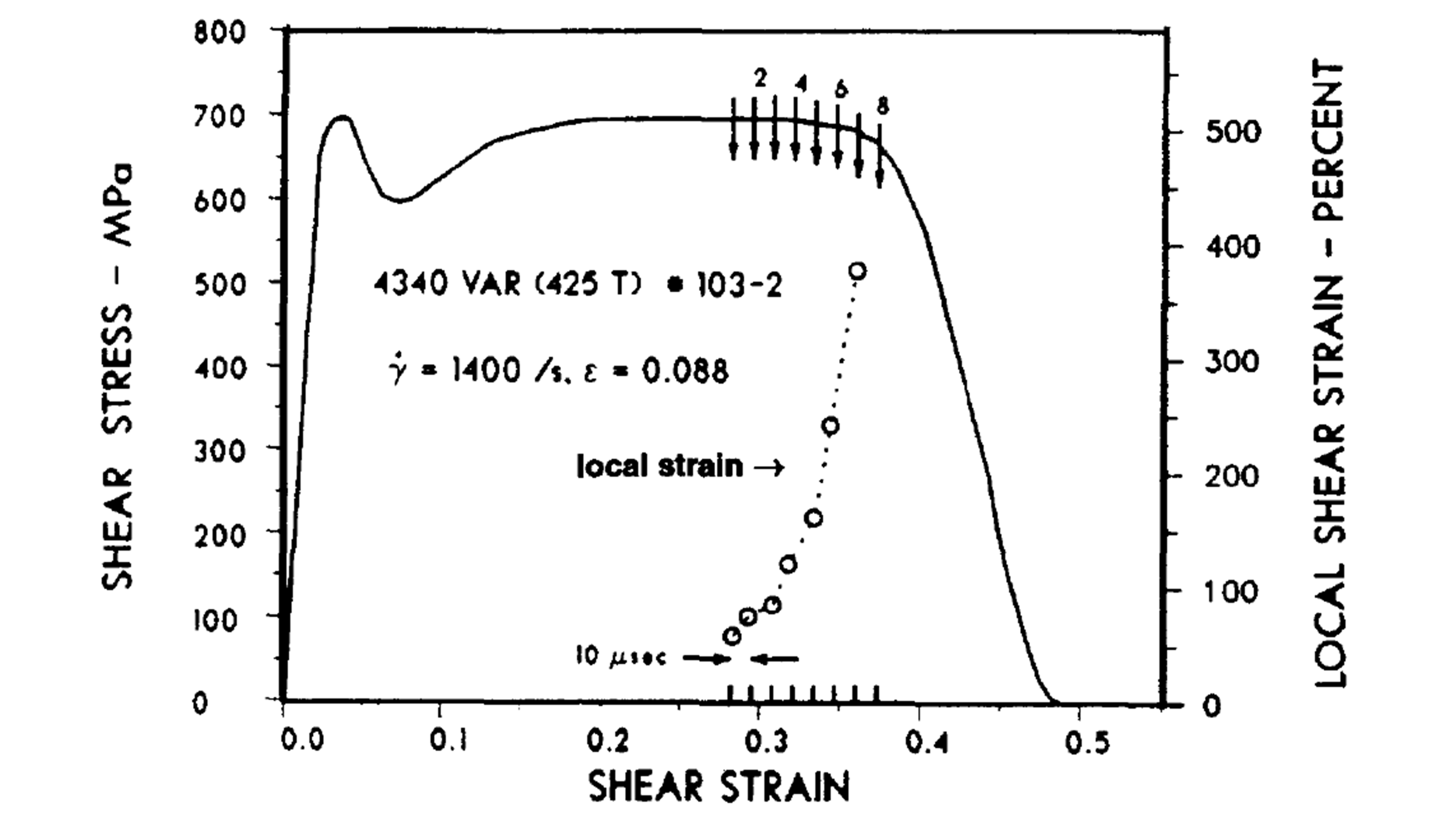}
  \caption{}
  \end{subfigure}
  
  \caption{(a) High-speed photographs (40 $\mu$s inter-frame time) of the grid patterns showing evolution of the plastic flow from a slightly inhomogeneous strain distribution (frames 1 and 2) to a well-defined shear band (frames 7 and 8) in high-speed torsion of I018 cold-rolled steel at a nominal strain rate of 1400 /s. The deformation stages referred to in  frames 1 through 8 are identified on the shear stress-strain curve shown in (b). Source: Ref.~\cite{marchand1988}.}
  \label{fig:highSpeedImages}
\end{figure}

With subsequent developments in high-speed imaging \cite{duffy1992,cho1990,cho1993,mgbokwere1994}, more detailed observations could be made of shear band evolution at a fixed position on the specimen at temporal resolution of $< 100\, \mu$s. Figure~\ref{fig:highSpeedImages} shows a sample high-speed photographic sequence, taken during band formation in dynamic torsion of 4340 steel at a strain rate of 1400 /s; the corresponding (nominal) stress-strain curve is also shown in the figure. The grid lines seen in frames 1-8 were initially straight and oriented parallel to the axis of the specimen (horizontal in figure); the slope of these lines at any position across the gage section therefore provides a measure of local shear strain. In Fig.~\ref{fig:highSpeedImages}(a), the strain distribution remains homogeneous across the specimen until a strain of $\sim 0.23$, corresponding to the maximum stress, see Fig.~\ref{fig:highSpeedImages}(b). This is followed by inhomogeneous flow, as deduced from the curved grid lines in frames 1-4. Continued straining results in the formation of a well-defined shear band plane (frame 5), followed by further localized \lq sliding\rq\ (plastic shear) along this plane under a continuously reducing load (frames 6-8). The end result is a highly localized deformation pattern, with strains inside the band typically being an order of magnitude higher than the nominally applied remote strain.

In stark contrast, quasi-static torsion experiments ($10^{-4}$ /s) on identical specimens showed no evidence of localization. These results provided the first direct experimental validation of Zener's hypothesis that the onset of inhomogeneous flow should coincide with the maximum in the stress-strain curve. They also demonstrated that fracture is not a pre-requisite for shear banding, but that, on the contrary, fracture may actually be preceded by localization even in very brittle metals \cite{duffy1992}. Additionally, the aforementioned temporal studies of stress, strain, and temperature histories during shear band formation under nominal simple shear conditions have also stimulated much interest in the mathematical modeling of shear banding; see Ref.~\cite{clifton1994}. In our view, the work of Duffy and co-workers deserve more critical study, and could provide a starting point if we are to fully understand shear band evolution quantitatively.

\begin{figure}
  \centering
\begin{subfigure}[b]{0.72\textwidth}
\centering
  \includegraphics[width=1\textwidth]{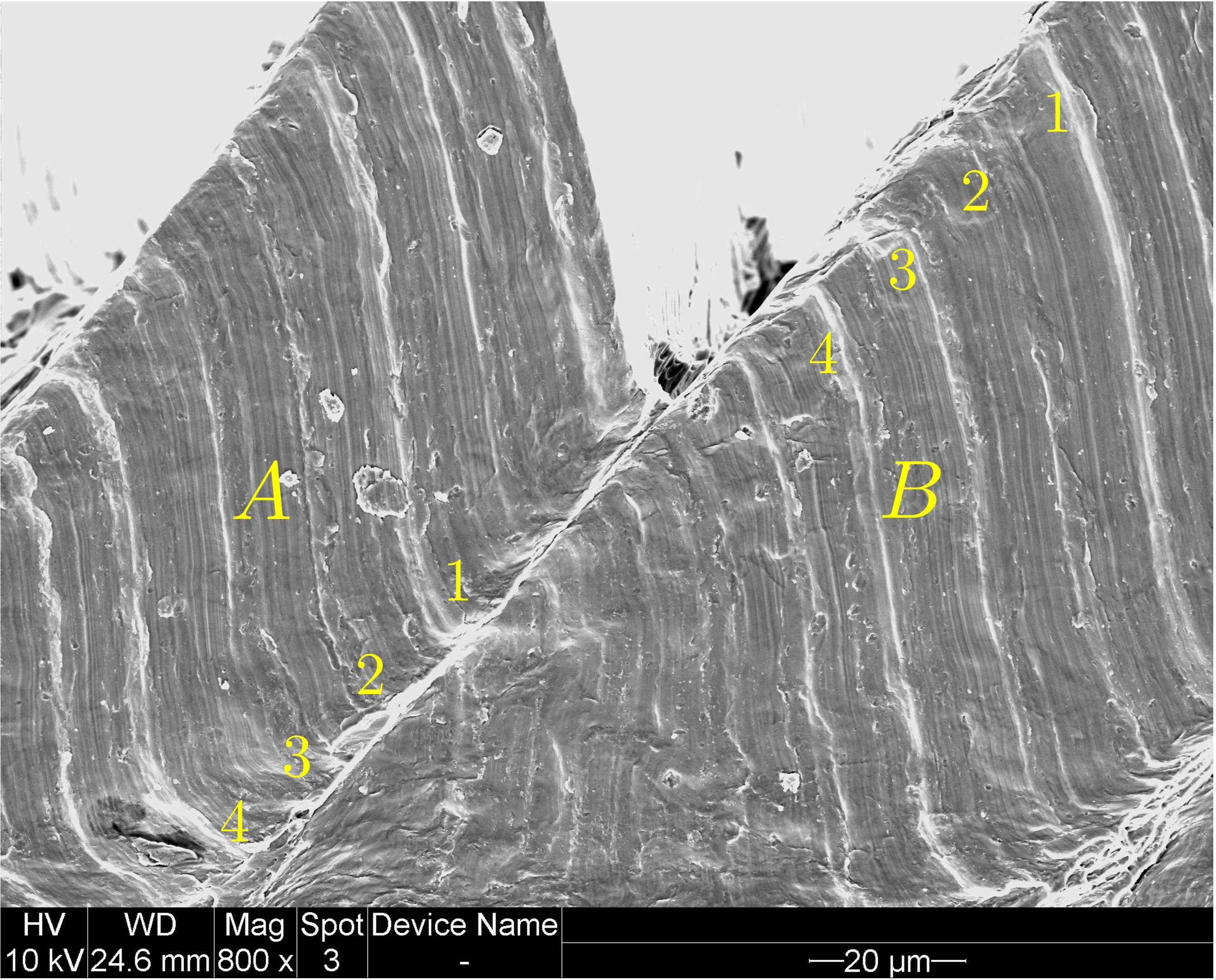}
  \caption{}
  \end{subfigure}

\begin{subfigure}[b]{0.52\textwidth}
\centering
  \includegraphics[width=1\textwidth]{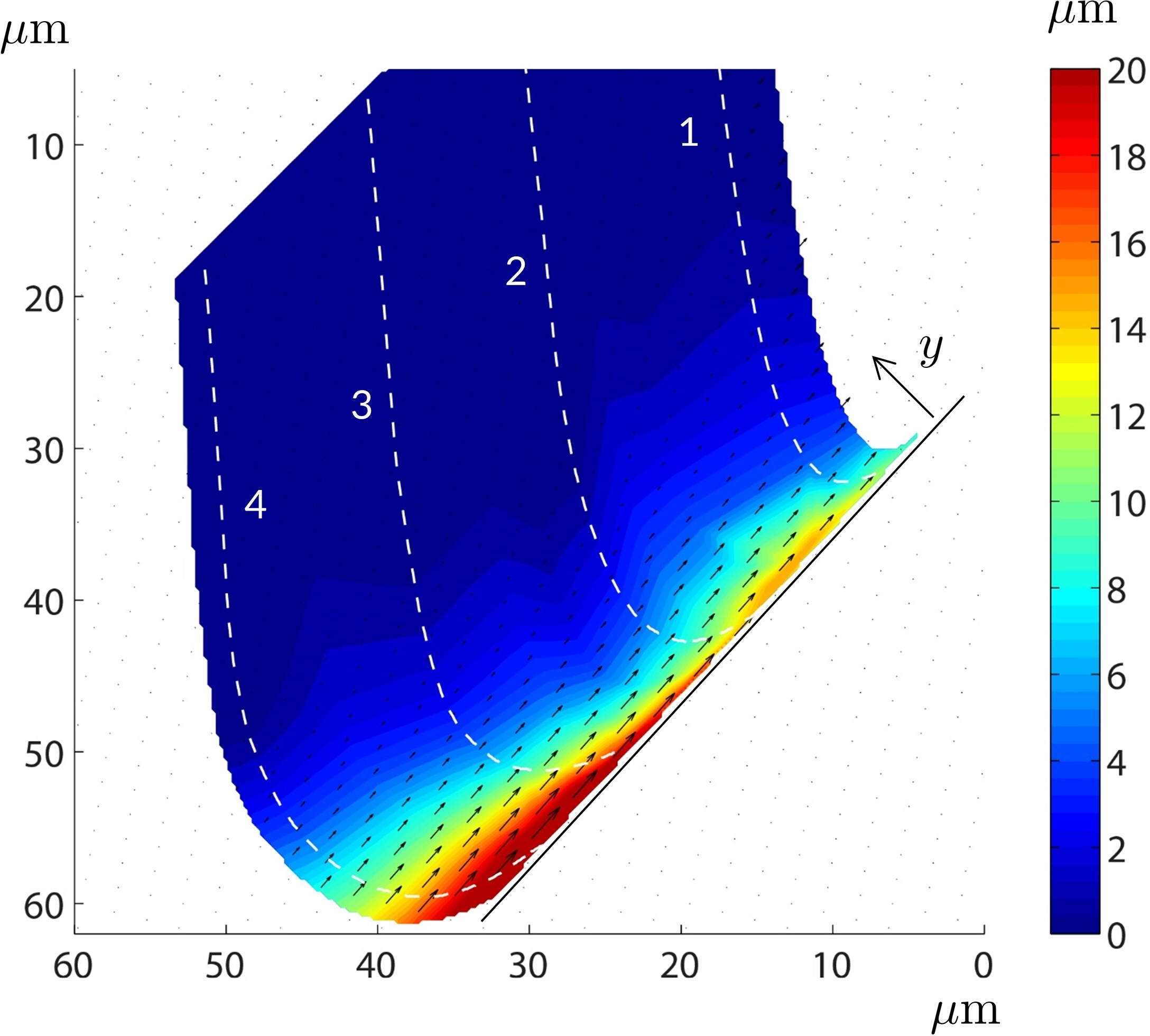}
  \caption{}
  \end{subfigure}
~
\begin{subfigure}[b]{0.44\textwidth}
\centering
  \includegraphics[width=1\textwidth]{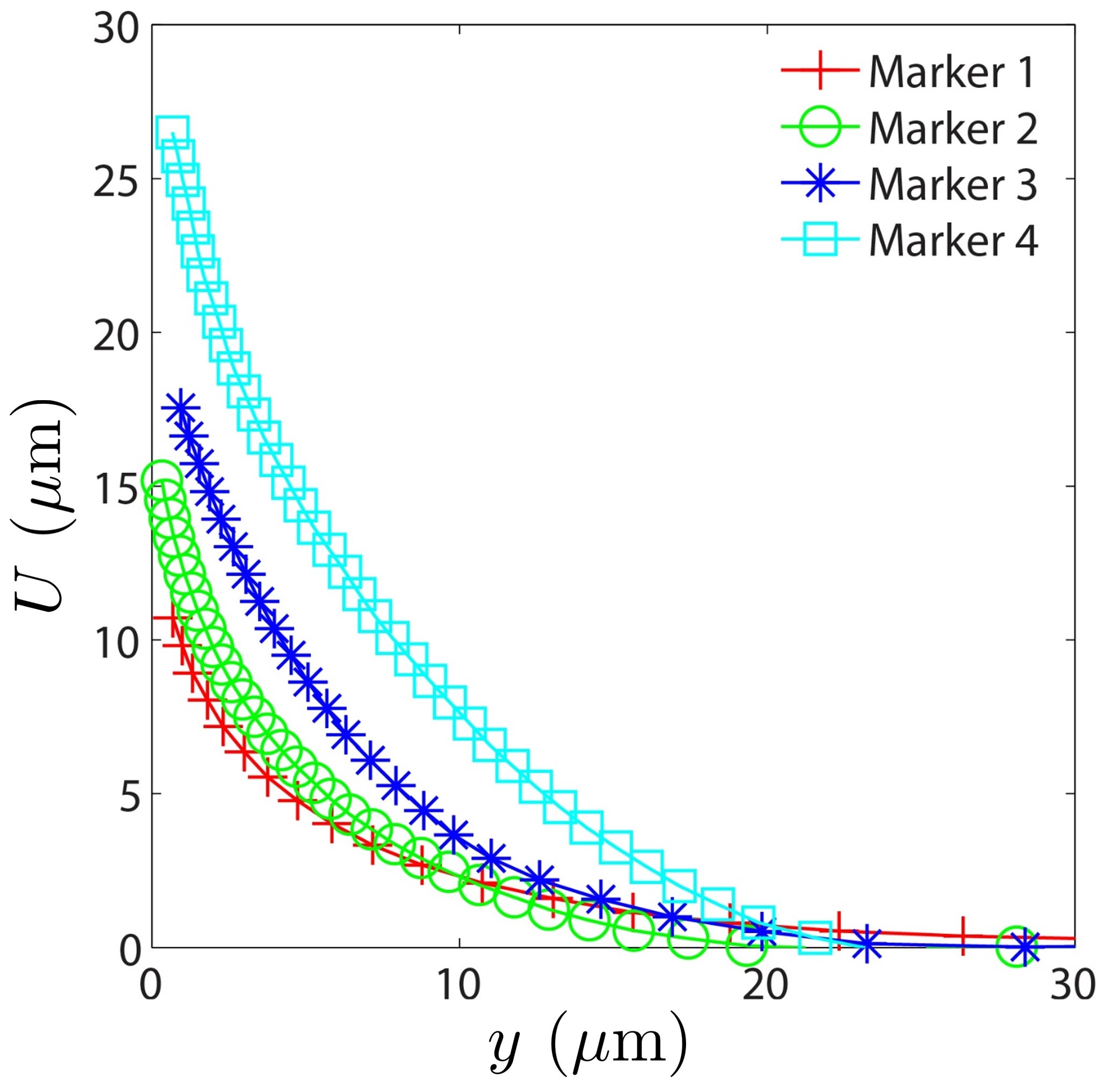}
  \caption{}
  \end{subfigure}
  
  \caption{(a) SEM image showing the morphology of markers around a shear band formed in a machined  Ti-6Al-4V chip. Four markers are labeled 1-4 to illustrate their relative positions on either side of the band. (b) and (c) show the full-field displacement field and individual displacement profiles computed based on the distribution of markers. Data only on one side (segment A) of the shear band are shown here. Steep displacement gradients close to the band center ($y = 0$) demonstrate very large shear strains $> 10$. The fluid boundary layer-like structure is also evident from the displacement field shown in (b).  Source: Ref.~\cite{sagapuram2018b}.}
  \label{fig:DSKVexpts}
\end{figure}

Even though these pioneering experimental studies represent a fundamental advance in the mechanistic understanding of shear band evolution, the  displacement/velocity profiles within (or in the immediate vicinity) of the narrow shear band could not be fully resolved in these studies. Specifically, the spatial resolution of the optical cinematography employed was not adequate for this purpose, except in the case of relatively thick bands $\sim 300$-$500\,\mu$m, as in ductile 1018 steel \cite{duffy1992}. It is only recently that the displacement field around a single  band has been more fully characterized at sub-$\mu$m resolution by Sagapuram and co-workers \cite{sagapuram2018a, sagapuram2018b, sagapuram2016}, combining \emph{in situ} imaging and \emph{ex situ} SEM analysis. By inscribing periodic scribe marks on the sample, the local deformation in the vicinity of a single shear band could be mapped at very high spatial resolution, see Fig.~\ref{fig:DSKVexpts}. A single band formed during machining of Ti-6Al-4V alloy is shown in Fig.~\ref{fig:DSKVexpts}(a), where inscribed markers appear as periodic striations in the image. Large shear displacement across the band are illustrated by relative positions of four markers (labeled 1-4) on either side of the band. Attention must be drawn at this juncture to the striking similarity between the marker profiles in Fig.~\ref{fig:DSKVexpts}(a) and grid lines in Duffy's torsion experiments (Fig.~\ref{fig:bandVicinityFlow}).

Full-field displacements may be readily calculated from micromarker displacements, see Fig.~\ref{fig:DSKVexpts}(b). Displacements are computed parallel to the shear band plane on one side (left) of the shear band. The corresponding (1D) displacement profiles $U(y)$ for four different markers (1-4) as a function of normal distance $y$ from the  band interface are shown in Fig.~\ref{fig:DSKVexpts}(c). $U(y)$ clearly shows a characteristic non-linear decay with $y$. Additionally, the severe localization of displacements in a narrow zone, as well as large displacement gradients (Figs.~\ref{fig:DSKVexpts}(b),~\ref{fig:DSKVexpts}(c)) adjacent to the shear band interface are also evident. In fact, these deformation fields quite closely resemble flow fields in shear of viscous fluids \cite{batchelor1967,oldroyd1947}. A small gradient in the displacement field is also seen along the length of the band (Fig.~\ref{fig:DSKVexpts}(c)), but this gradient is about an order of magnitude smaller than the gradient normal to the band. This may also be shown from a simple scaling analysis: $\partial U/\partial x$ must scale $V_S/L$, with $V_S$ being the relative sliding velocity across the band and $L$ the band length, while $\partial U/\partial y \sim V_S/\delta$, $\delta$ being the band thickness. Since $\delta \ll L$, the gradient normal to the band is clearly much larger.

While this micromarker technique has been extended to capture  strain-rate effects on shear band displacement fields \cite{sagapuram2018a}, it is still subject to the constraint that it cannot fully resolve the temporal evolution of the local displacement/strain fields in the vicinity of a nucleated shear band.



Finally, the fact that in machining, shear bands initiate at the tool tip,  propagate towards the free surface, followed by localized material flow/sliding (Fig.~\ref{fig:sb_initiation}) is quite consistent with observations in  dynamic torsion \cite{marchand1988} and plate impact experiments \cite{wingrove1973}. This only serves to show that the continuum-scale phenomenological features of shear band dynamics, involving distinct phases of initiation and growth (strain development), are universal to all metal systems and loading configurations. 

\subsection{Shear band viscosity and  boundary layer analysis} \label{sec:boundarylayer}

In contrast to band nucleation, the growth phase appears to have little consideration for specific microstructural details of the material. This is not  entirely impossible to imagine, given the severe shear strains ($>10$) within the band and the short time scales in which they develop. This, therefore, suggests the possibility of understanding plastic flow evolution around the band based purely based on continuum considerations.

\begin{figure}
  \centering

  \includegraphics[width=0.8\textwidth]{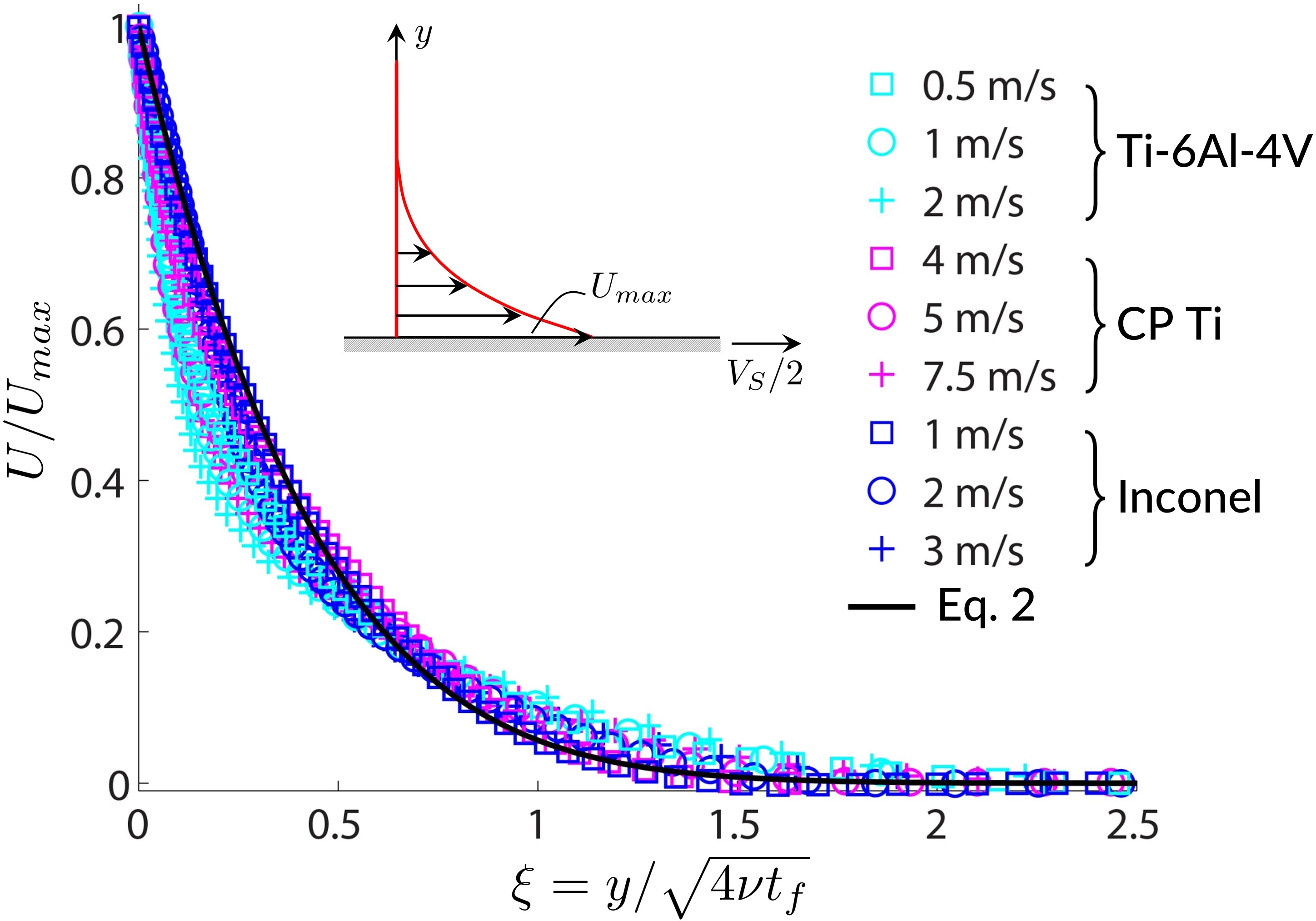}

  \caption{ Normalized shear band displacement plots for three metals: Ti-6Al-4V, commercially pure (CP) Ti and Inconel 718 across different strain rate conditions \cite{sagapuram2018b}. The shear band displacement profiles were determined using the marker method and fitted using the $\nu$ parameter. The profiles for all three materials are seen to be very closely spaced and coincide with the theoretical profile (black solid line) predicted by the Bingham model (Eq.~\ref{eqn:theoretical}).}
  \label{fig:DSKVviscosity}
\end{figure}

To this end, Fig.~\ref{fig:DSKVviscosity}  shows the shear band displacement data for three microstructurally distinct materials (pure Ti, Ti-6Al-4V alloy and Ni-based superalloy), obtained using inscribed micromarkers. Displacement profiles of shear bands produced under different imposed nominal strain rates are plotted together in the same figure after rescaling. The vertical axis shows the normalized displacement $U(y)/U_{max}$, where  $U_{max}$ is the maximum displacement at the band center, $y = 0$ (see Fig~.~\ref{fig:DSKVexpts},~\ref{fig:DSKVviscosity} (inset)). The horizontal axis shows the non-dimensionalized distance $\xi = y/\sqrt{4 \nu t_f}$ where  $t_f = 2 U_{max}/V_S$ is the time taken for the band to fully develop and  $V_S$ is the relative (sliding) velocity across the shear band. Here  $\nu$ is a parameter that may be suitably chosen so that the displacement data for all three materials, and across strain rates, collapse onto a master curve (Fig.~\ref{fig:DSKVviscosity}). This is clearly indicative of a common band growth mechanism across different materials and rates. Furthermore,  the form of $\xi$ suggests a diffusion-type description of the flow, with $\nu$ playing the role of a diffusion coefficient.

In fact, in view of the above observations, \emph{viz.} diffusion-like flow kinematics, lateral growth of band thickness with time, and negligible variation along the band length direction,  one could easily envision modeling the strain-intensive band growth phase as a one-dimensional velocity-driven fluid flow problem. Treating the material in the immediate vicinity of the shear band as a linear rate-dependent solid with a yield stress $\tau_0$ (the Bingham solid) and imposing no-slip boundary condition at the band interface ($y = 0$), the displacement field in the band vicinity can be theoretically derived as \cite{batchelor1967}:  
\begin{equation}
  \label{eqn:theoretical}
  \frac{U}{U_{max}} = 1 - 2\xi^2 \erfc(\xi) + \erf(\xi) + \frac{2\xi}{\sqrt{\pi}} \exp(-\xi^2)
\end{equation}
where $\xi = y/\sqrt{4 \nu t_f}$, with $\nu$ now being the kinematic viscosity of the Bingham material and $t_f$ is as before. It should be noted that the no-slip boundary condition is justified by the fact that displacements are continuous across the band interface, without any signs of fracture. Secondly, since the strains surrounding the band well-exceed the yield strain of the material,  the viscoplastic Bingham model kinematically reduces to a Newtonian fluid model for the velocity/displacement fields \cite{pascal1989}.  

 From Eq.~\ref{eqn:theoretical}, it can be seen that the theoretical displacement curve also scales as a function of $\xi$, so that the fitting parameter $\nu$ from Fig.~\ref{fig:DSKVviscosity} may now be interpreted as  kinematic viscosity for material flow near the band interface. From Fig.~\ref{fig:DSKVviscosity}, it is at once clear that the theoretical curve, which is shown as a solid black curve, nearly coincides with the collapsed experimental data, suggesting that the band growth is indeed dominated by viscosity effects.  

 Furthermore, the shear band (dynamic) viscosity $\mu = \rho\nu$ ($\rho$ is density) obtained from the displacement profiles is found to be very small for all three materials, of the order of a few mPa$\cdot$s. These values are typical of liquid metal viscosities and raise interesting questions about their origin \cite{sagapuram2018b}. Incidentally,  metal interfaces at the microstructure level such as  slip bands \cite{zener1948b} and  grain boundaries \cite{giannuzzi1985} are also known to exhibit this type of viscous behavior. Nonetheless, the fact that three very different material systems appear to somehow show the same continuum-scale flow behaviour reaffirms the fact that the details of deformation are largely microstructure independent. 

In addition to describing flow kinematics, the Bingham model also provides useful information about the evolution of stresses. From the Bingham constitutive relation, temporal evolution of the shear stress at the band interface can be shown to follow $  \tau \propto  \dfrac{\mu V_S}{\sqrt{4\pi \nu t}}$, where $\mu$ and $\nu$ are, as before, the dynamic and kinematic viscosities and the time $0\leq t\leq t_f$ is measured from the onset of sliding. 
This $1/\sqrt{t}$ dependence of the shear stress is in qualitative agreement with the rapid stress collapse that is experimentally observed following band initiation \cite{marchand1988,yadav2020a}. Subsequent to this, material surrounding the band flows at a nearly constant stress $\tau_0$, which, from force measurements, is determined to be almost equal to the shear yield strength of the material \cite{sagapuram2018b}.

Based on these displacement measurements and the Bingham-type flow behavior, one can \emph{a posteriori} ascertain that plastic flow in the vicinity of a single band is, in fact, quite like a boundary layer phenomenon. The necessary conditions for flow of a Bingham fluid to localize in the form of a boundary layer near a narrow zone are the following \cite{oldroyd1947}. Firstly, that the Bingham number be much larger than the  Reynolds number, $Bi \gg Re$, and secondly, that the band thickness to maximum displacement ratio obeys $\left(\delta/ U_{max} \right)^3 Bi \sim 1$ and $\left(\delta/ U_{max} \right)^2 Re \ll 1$. For the present case, the Bingham number $Bi = \tau_0 U_{max}/(\mu V_0) \sim 10^6$, whereas the Reynolds number $Re = V_S U_{max}/\nu \sim 20$ so that all three conditions are simultaneously satisfied. The boundary layer characteristics are also quite evident from the experimental observations. Just as in a viscous boundary layer in fluid flow, plastic deformation here remains confined to a very narrow zone near the nucleated band interface (Figs.~\ref{fig:bandVicinityFlow} and \ref{fig:DSKVexpts}(b)). These observations together suggest that shear banding belongs to the broad class of laminar boundary layer problems. While attempts to model shear bands using boundary layer analyses are not new \cite{gioia1996, dilellio1997}, it is only quite recently that quantitative comparisons have been made between flow field measurements and model predictions \cite{sagapuram2018b,yadav2020}.

Lastly, given the material-agnostic description of strain fields during band development, several open questions remain.

\begin{enumerate}

\item What is the origin of a Bingham-type flow rule, and under what conditions is the linear rate dependence valid? Can this be explained using a material-independent microscopic model from a dislocation mechanics standpoint?

\item \markthis{Most constitutive laws contain non-linear strain rate dependence of the flow stress and temperature-dependent viscosity, in addition to strain hardening term. So the linear rate-dependent rule with constant viscosity may prove to be limiting in some cases, and certainly warrants further investigation. For this, we highlight the recent work of Tzavaras \cite{tzavaras1986,tzavaras1992} who studied the asymptotic behavior of incompressible non-Newtonian fluids with temperature-dependent viscosity in rectilinear shear flows (under both prescribed velocities and prescribed tractions). An important consequence of this investigation to is the existence of  a unique solution that converges to steady-state uniform shear flow as $t \to \infty$, given a monotonic temperature dependence of the viscosity. These observations are qualitatively consistent with the above Bingham model as well as recently reported time-resolved displacement measurements around an evolving shear band \cite{yadav2020}.}

\item What is the origin of very small apparent band viscosity? Recent studies  suggest phonon drag on mobile dislocations at very high strain rates ($> 10^4$ /s) as a plausible microscopic mechanism contributing to the viscosity\cite{sagapuram2018a,sagapuram2018b}.

\item Lastly,  the fact that a boundary layer-type description applies equally well to different material systems raises a question about the physical significance of the Bingham and Reynolds numbers in the present situation. Are they reflective of underlying physical principles on the microscale? 
\end{enumerate}

Answers to these questions hinge on detailed  investigation of other analytical and semi-numerical techniques coupled with experiments focused on plastic flow evolution around a single shear band.

\subsection{Temperature rise, thermal diffusion and shear band width}

The thermal fields surrounding a developing shear band are perhaps even less understood compared to mechanical/kinematic fields (displacement field, strain, velocity, etc.) discussed in the earlier section. Even though it is generally agreed that shear bands formed under high strain rates are subjected to remarkable thermal excursions (temperature rise up to $\sim 1000$ $^\circ$C in a few $\mu$s), direct spatiotemporal measurements of the temperature profiles in the vicinity of single bands are few and far in between. The earliest attempt to characterize the local temperature rise was by Tresca \cite{tresca1878} where the effect of local heating was revealed ingeniously by coating the lateral sides of the lead specimen with a thin layer of wax. Upon the deformation, it was found that wax had preferentially melted in the regions of maximum shear, outlining the shear band pattern in the form of an \lq X\rq . A related, more recent, technique uses fusible tin coating to characterize the heat production at shear bands in metallic glasses \cite{lewandowski2006}.

In this context,  we must also draw attention to Basinski's experiments \cite{basinski1957, basinski1960} on the  discontinuous load reduction (shear bands?) in tensile testing of aluminum at liquid helium temperatures (4 K). For measuring the temperature rise during load reduction, a superconducting niobium wire was placed within the tensile specimen. Concurrent with the reduction, a portion of the niobium wire was converted from the superconducting state to the normal state, showing that the temperature of that part of the specimen rose at least above 9 K. While the magnitude of this temperature rise in itself was small, this result was important in showing that the discontinuities in the stress-strain curve and associated strain localization during plastic deformation of metals at very low temperatures arise from an instability that is also of thermal origin. Basinski's studies highlighted, perhaps for the first time, that it is the rate of temperature rise with strain and relative change in the flow stress with respect to deformation variables (strain, strain rate and temperature) that govern the instability onset. This was quite in keeping with Zener's original formulation \cite{ZenerHollomon_JApplPhys_1944}.

Post-mortem metallurgical observations of the shear band microstructure have been also widely used to deduce the shear band temperature by other researchers. For example, the fact that shear bands in steels often appear as featureless \lq white\rq\ streaks under a nital etch has been taken as evidence that the material has locally been heated to at least the austenitic phase transformation temperature ($\sim 750^\circ$C) which, upon rapid cooling by surrounding material, forms the martensite phase \cite{Rogers_AnnRevMatSci_1979}. However, this type of interpretation should be considered carefully, since the featureless appearance of white layers in steels can also be a consequence of extremely small size of the grains/crystals therein \cite{akcan2002}. Even when we have direct evidence of phase transformation (for example, from TEM analysis \cite{glenn1971,wingrove1971}), caution should be exercised in inferring the actual temperatures since rapid diffusion-based phase transformation in a few $\mu$s (typical shear band times) necessitates that the temperature be considerably higher than the equilibrium transformation temperature \cite{archard1988}.

It is  noteworthy that, as with displacement/strain field measurements, an important advance in tracking the temperature rise during shear band formation was also made by Duffy's group at Brown University \cite{costin1979,hartley1987}, using non-contact infrared radiation thermography. In the first such study \cite{hartley1987}, a linear array of ten indium-antimonide detectors was used to track temperature values at ten neighboring points across a single  shear band in 1020 steel under dynamic torsion; this enabled temperature profile across the shear band to be measured as a function of time ($\sim 1\,\mu$s time interval). Several improvements to the measurement technique in terms of number of detectors, temporal resolution and measurement spacing have been made in subsequent studies \cite{duffy1992,mgbokwere1994,liao1998}, although the spot size (17-35 $\mu$m) has remained larger than the typical shear band thicknesses. Therefore, the recorded temperature rise in the band ($\sim 400$-$500^\circ$C in steels and Ti-6Al-4V alloy \cite{duffy1992,mgbokwere1994}) likely represents a lower-bound on the actual temperature, since the the measurements are averaged over both the \lq hot\rq\ shear band area and adjacent cooler, low-strain regions. Notwithstanding this limitation, these studies have elucidated key thermal features of shear bands, \emph{viz.} large thermal gradients across the band thickness with peak temperature occurring at the band center; a temperature zone that is significantly wider (5-10 X) than the shear band thickness; a sharp increase in the temperature that is coincident with the stress collapse within the band; relatively low temperatures ($< 200^\circ$C) at the point of shear band initiation; and extremely high heating/cooling rates of the order of $10^6$ to $10^7$ $^\circ$C/s associated with band formation and arrest. Note that sharp temperature rise during the stage of stress collapse---equivalently, shear band growth phase---is consistent with the concomitant large  strain accumulation ($> 10$). It is perhaps also worth noting that the relatively small temperature rise at the time of shear band nucleation has led some researchers \cite{rittel2008,guo2019} to propose that the temperature rise is not the only trigger for band formation. As we pointed out earlier, it is the relative change of flow stress with respect to temperature, strain and strain rate that should be taken into account when analyzing the shear band initiation, and not their absolute values.

While the above time-resolved measurements have provided insight into the dynamics of temperature-related events, accurate measurement of the temperature \lq field\rq\ in and around the  band was not possible since a linear array of detectors can measure temperatures only along a series of discrete points across the band. Coarse full-field measurements of the temperature over a 2-dimensional area covering the shear band have been made by Guduru \emph{et al.} \cite{guduru2001}. 
While these full-field measurements could be made at a rate of 1 MHz, the spatial resolution (100 $\mu$m) of this technique was still significantly larger than the typical band thickness. A later study \cite{ranc2008}, which utilized an intensified CCD camera with a GaAs photocathode and short exposure time of 10 $\mu$s, has by far provided the highest resolution (2 $\mu$m) temperature field data to date. However, the large refresh time of the CCD sensor  (10 ms) meant that only a single temperature snapshot was possible. Nonetheless, the resulting data showed non-homogeneous temperature distribution along the length of the band, characterized by \lq hot spots\rq . The question of whether these hot spots are a general phenomenon, as well as the role they play in the overall shear localization-to-fracture transition are far from clear.

Concerning the shear band width, it is customary in most continuum analyses to associate this length scale with the thermal length scale resulting from the thermal diffusivity parameter \cite{dodd2012}. A simple analysis of the characteristic length scales associated with thermal diffusion is pertinent at this point. If we assume that each shear band is an infinitesimally thin, continuous planar heat source of constant strength $Q$ acting throughout the sliding/growth phase, then the (one-dimensional) temperature field around the band can be given as \cite{carslaw1959}:
\begin{equation}
  T - T_0 = \frac{Q}{\kappa \rho C} \left[\sqrt{\frac{\kappa t}{\pi}} \exp\left(\frac{-y^2}{4\kappa t}\right) - \frac{y}{2} \erfc\left(\frac{y}{\sqrt{4 \kappa t}}\right)\right]
\end{equation}
where $T_0$ is the temperature just before band nucleation (at $t=0$) and $y$ is, as before, the perpendicular distance from the band interface. Other variables are the thermal diffusivity $\kappa$, material density $\rho$ and heat capacity $C$, and time $t$. For shear bands,  $t_f \sim 10$-$100\,\mu$s, so that the  thermal boundary layer thickness (i.e.,  zone in which temperature rise is at least 1\% that at the center) is $\sim 100\,\mu$m. While this value  matches reasonably well in ductile metals such as copper and 1018 steel, it is actually about an order of magnitude larger than the typical shear band thicknesses found in most materials. More careful analyses \cite{batra1992} have also shown that the observed shear band widths do not seem to exhibit a direct relationship with the thermal conductivity of the material. \markthis{This has motivated alternative approaches to explain shear band width evolution post-initiation, most notable of which is the use of strain gradient plasticity. Here an internal length scale is introduced into the formulation by incorporating higher order strain gradients into the  yield condition or constitutive equation for the flow stress. In the context of shear banding, this approach was first introduced by Aifantis \cite{aifantis1987,zbib1989} who showed that the width of the band is proportional to the gradient coefficient. However, we must note that strain gradient theories typically work best when the gradients (not necessarily absolute strains) are very large; in the context of materials or deformation processing, both strains and strain gradients are commonly very high. As an example, while strain-gradient models have helped explain qualitative scaling relationships between band width and applied strain rate, the predicted band widths are typically in the order of mm, much too large for exact comparison with experimental observations (few tens of $\mu$m).}

Going back to the Bingham fluid analogy of the previous section, the viscosity parameter provides an alternative way to define the shear band width in terms of momentum diffusion. For example,  analogous to the thermal diffusion length $\sqrt{\kappa t_f}$,  the momentum diffusion length $\sqrt{\nu t_f}$ may be obtained from the kinematic model discussed earlier (Sec.~\ref{sec:boundarylayer}). This shows that the corresponding Prandtl number $Pr = \nu/\kappa$ for shear band flow is indeed very small ($\ll 1$). These results seem to suggest that, at least in the absence of strain gradient effects, momentum diffusion is the
dominant factor determining the shear band width. For the interplay between thermal and momentum diffusion and their effect on shear band development, we point to the work by Grady and co-workers \cite{grady1987, grady1992}, who  made a number of important predictions about shear band spacing and evolution, and showed that the width of the band is determined by a balance between these two diffusive mechanisms \cite{grady1987}. The analysis presented above is complementary to these studies in that momentum and thermal processes are treated individually.  

To correlate the experimental data with analytical/computational models for post shear band initiation phases, one would ideally like to investigate the time-evolution of local shear band displacement (strain), velocity (strain rate) and temperature fields under controlled and wide-ranging loading  conditions (e.g., strain rate, stress state) and material systems. 
Very little, if any, work has been done along this direction. With recent developments in high-speed imaging systems (both in the visible and infrared range) and digital image correlation techniques, opportunities now exist to simultaneously map kinematic and thermal fields around single shear bands at high spatial and temporal resolution, and provide ample opportunity for scientific study.

\section{Transition from banding to fracture}
\label{sec:fractureTransition}

The basic motivation for studying the onset and dynamics of shear banding remains that it is very often a precursor to catastrophic dynamic fracture \cite{latanision1987}. In general, shear bands retain their structure and remain contiguous with the surrounding material until the initiation of a crack. The entire process represents a double transition to fracture---first a bifurcation from homogeneous deformation to strain localization, followed by a second transition to fracture and physical material separation. The dynamics of this second transition has some very interesting features that have been the subject of significant experimental and theoretical investigation.

\begin{figure}
  \centering

  \includegraphics[width=0.8\textwidth]{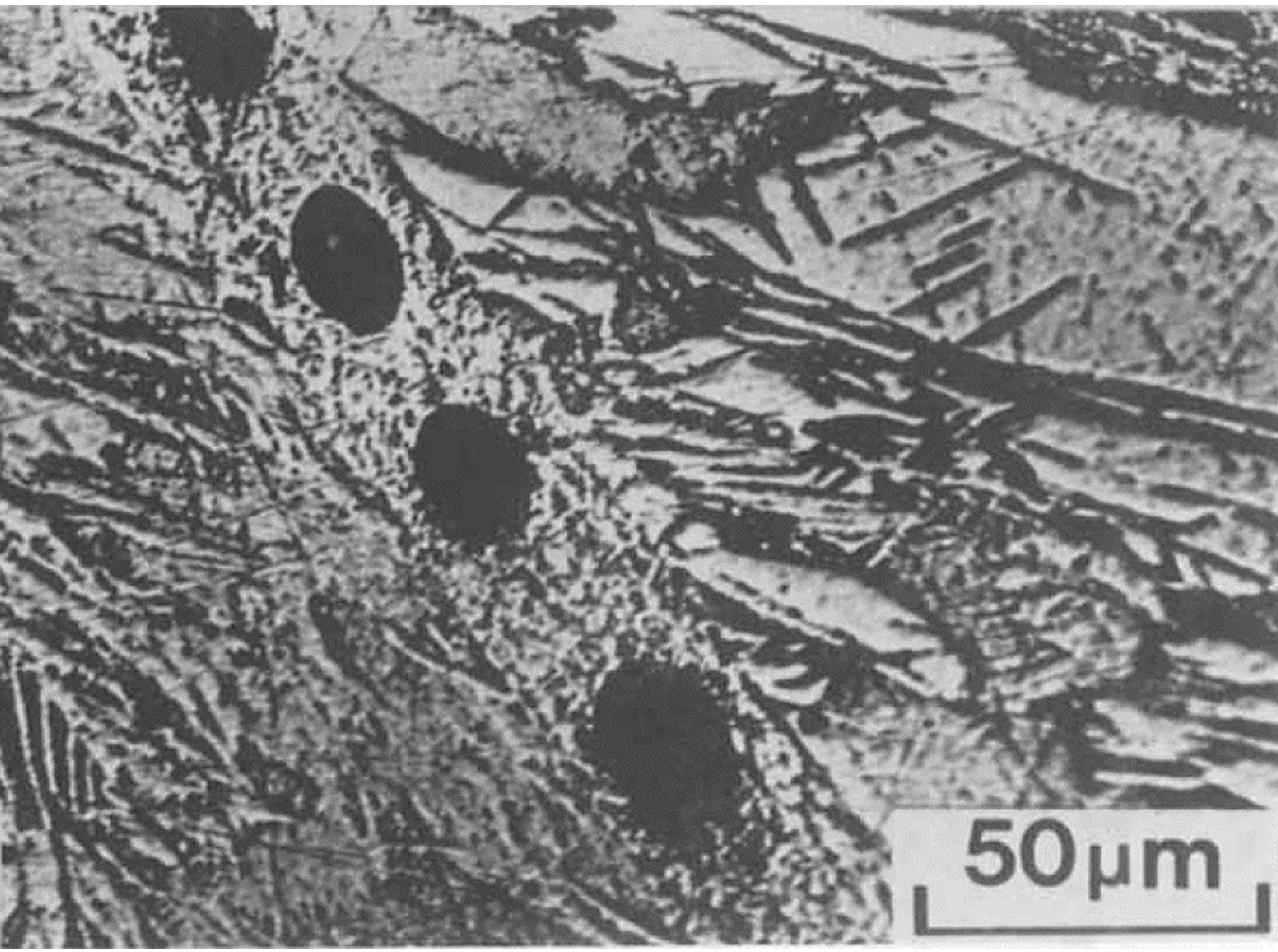}

  \caption{Micrograph showing a shear band with spherical cavities formed in an impact crater in titanium after impacting with a flat ended steel rod  at 330 m/s \cite{winter1975a}. }
  \label{fig:fracture}
\end{figure}

A configuration in which the transition is often easily observed is in projectile impact \cite{StockThompson_1970, TimothyHutchings_1985, timothy1985, wingrove1973, Winter_1975}. While the precise mechanism for the appearance of macroscopically identifiable cracks is yet unclear, detailed experiments have pointed to several interesting possibilities. That fracture does occur is commonly appreciated, even by the erosion and wear community, where banding-mediated material separation is a well-accepted mechanism for material removal \cite{winter1975a, ShewmonSundararajan_1983, ChristmanShewmon_1979}. The usual sequence of events during projectile impact and the associated deformation is as follows \cite{timothy1985,wingrove1973}. Firstly, as the impact begins and in the early stages, a largely compressive stress state exists ahead of the projectile, making it conducive for the initiation and development of adiabatic shear bands. Secondly, when the strains inside the bands cross some critical value, voids are nucleated spontaneously, see Fig.~\ref{fig:fracture}. Finally, at the later stages of impact, a tensile unloading wave usually passes through the previously strained region, and these aforementioned voids grow in size, leading to the appearance of large macroscopic cracks. In fact, some authors have also proposed the homogeneous flow-to-banding-to-fracture route as a general mechanism for ductile failure in metals \cite{chin1964}.

For Ti-6Al-4V, an alloy well-known for its propensity to deform via shear banding, the formation of microvoids within shear bands at a critical strain value have indeed been directly observed \cite{timothy1985}.  Craters formed by the impact of hardened steel balls on Ti-6Al-4V samples have shown distinct bands in the sub-surface where both void formation and coalescence were observed using post-mortem metallographic analysis. The shape of these microvoids determined the morphology of the resulting fracture surfaces. Unfortunately, very little quantitative data appears to exist beyond such phenomenological observations.

In fact, the problem of fracture initiation is more complex than this picture suggests, as shown by the following calculation. Based on the displacement measurements in the vicinity of single bands (Sec.~\ref{sec:shearBandDevelopment}), typical shear band strains measured in Ti-6Al-4V are $\sim 10$. From Fig.~\ref{fig:DSKVexpts},  it is clear  that even at these large strains, the material on either side of the band remains continuous. Furthermore, the Bingham fluid model provides an estimate of the shear stress within the band as being nearly half the tensile yield stress, $\sim 350$ MPa \cite{sagapuram2018b}. Using these results, one can now back calculate the maximum allowed void size inside the band so that the material is still continuous macroscopically. If the fracture toughness of Ti-6Al-4V is taken to be 90 MPa$\cdot$m$^{1/2}$ (a conservative estimate), then this gives a maximum void size of $ \sim 4$ cm!  This is clearly too large---the corresponding voids would be larger than the band itself! This likely indicates that the mechanisms of collapse in the present case of mode II loading with large pre-strains are very different from the conventional mode I measurement usually used to determine fracture toughness. So the simple picture of coalescing voids inside the band somehow resulting in crack growth is clearly lacking in quantitative detail.  


An interesting alternative hypothesis in this regard was proposed by Teng, Wierzbicki and Couque \cite{TengWierzbicki_2007}. They showed using detailed numerical simulations that the void nucleation was governed, not by a critical strain criterion, but instead by local periodic occurrence of \lq hot spots\rq\ within a shear band.  Further, post-banding fracture was seen in their simulations only with the use of the Bao-Wierzbicki fracture locus \cite{BaoWierzbicki_2004}. Transition to fracture itself occurred by the growth of several microvoids (somewhat similar to Fig.~\ref{fig:fracture}) that coincided with local temperature hot spots and not by a single crack growth under a tensile unloading wave as proposed previously. It is interesting to consider what role these hot spots have in the formation of  white-etching regions  that are often observed inside  shear bands \cite{giovanola1988a} and which are indicative (if not conclusive) of a phase transformation.  It is possible that the development of hot spots could play a major role in the formation of (brittle) white-etching regions and thereby influence the subsequent development of voids.

If the problem of how cracks originate within the shear bands is, for the moment, kept aside, the nature of the fracture evolution and the role of corresponding deformation geometry also presents interesting questions. The complementary experiments of Kalthoff and Winkler \cite{KalthoffWinkler_1988} and of Zhou \emph{et al}. \cite{zhou1996} showed a remarkable transition from outright brittle fracture to shear-banding driven fracture (and vice-versa) as a function of projectile impact velocity and the deformation geometry (single vs. double edge notched plate). These experiments show that the effect of loading geometry and strain rate are central in determining how and when shear band-induced fracture occurs.

It is important to also point out that  a tensile loading wave (e.g., due to post-impact recovery), or even any form of tensile loading post band formation, is not really essential for voids to grow inside the shear band \cite{TorkiBenzerga_2018}. Even when the loading within the band is predominantly of shear, pre-nucleated voids can rotate and enlarge leading to collapse between neighbouring voids. This mechanism could be prevalent in non-impact type loading configurations, e.g., machining \cite{vyas1999}, and can possibly also explain highly elongated dimple-type fracture surface morphologies.

A practically significant implication of the shear band-induced fracture process is in the dynamic blanking of thin plates using a high-speed punch \cite{dodd1983,stock1971}, bringing us back full-circle to Zener and Holloman's original work \cite{ZenerHollomon_JApplPhys_1944}. The force needed to blank (punch a through hole) in a unit thickness of material is found to be much larger when the blanking is done quasi-statically than if it is done at high strain rate. This, somewhat counter-intuitive observation, is explained by the nature of plastic flow in the two cases. During quasi-static deformation the plastic zone size is quite large leading to significant energy dissipation over a large volume, whereas under high rates, deformation is localized to a thin band. However, the details of how localization transitions to fracture play a crucial role in determining the final surface quality and blank integrity \cite{BalendraTravis_1970}. At moderately high punch velocities ($\sim 25$ ft/s for steels), the localized deformation in the band appears to transition to a uniform fracture mode leading to a clean separation between the blank and the stock. But when the applied velocity is increased much more, defects start to appear at the blank edges, the most notable of which are the so-called \lq dishing\rq\ and \lq doming\rq\ defects \cite{davies1965}. This clearly indicates that strain-rate and temperature effects are not insignificant in the transition to fracture.

\section{Shear band control and some practical applications}
\label{sec:applications} 

Given the usually adverse consequences of shear banding and its relation to material failure in many facets of materials and deformation processing, a natural question to ask would be: Can we control shear bands? By \lq control\rq , we mean here the suppression of the highly non-homogeneous large strain flow that follows band initiation. As we have discussed up to this point, much scientific effort has been expended to develop a mechanistic understanding of shear band phenomena so that one may design materials that are less prone to shear banding in a particular application. On the other hand, the possibility of directly effecting shear band control/suppression has received very little attention in spite of potential technological benefits. In our view, even though many theoretical issues pertaining to band initiation and propagation remain (to be discussed in Sec.~\ref{sec:discussion}), much productive work can be done in the area of  arresting ensuing localized flow and associated fracture, irrespective of scientific details. This is in fact in line with Zener's view on how science should proceed \cite{zener_interview}:

\begin{displayquote}
  (Scientific) Work can proceed at different levels, one might say of sophistication, and you do not want to wait until what you might call the most basic problems are understood before you start thinking on the larger scale where you are not as concerned with the individual molecules.
\end{displayquote}

\begin{figure}
  \centering
\begin{subfigure}[b]{1\textwidth}
\centering
  \includegraphics[width=1\textwidth]{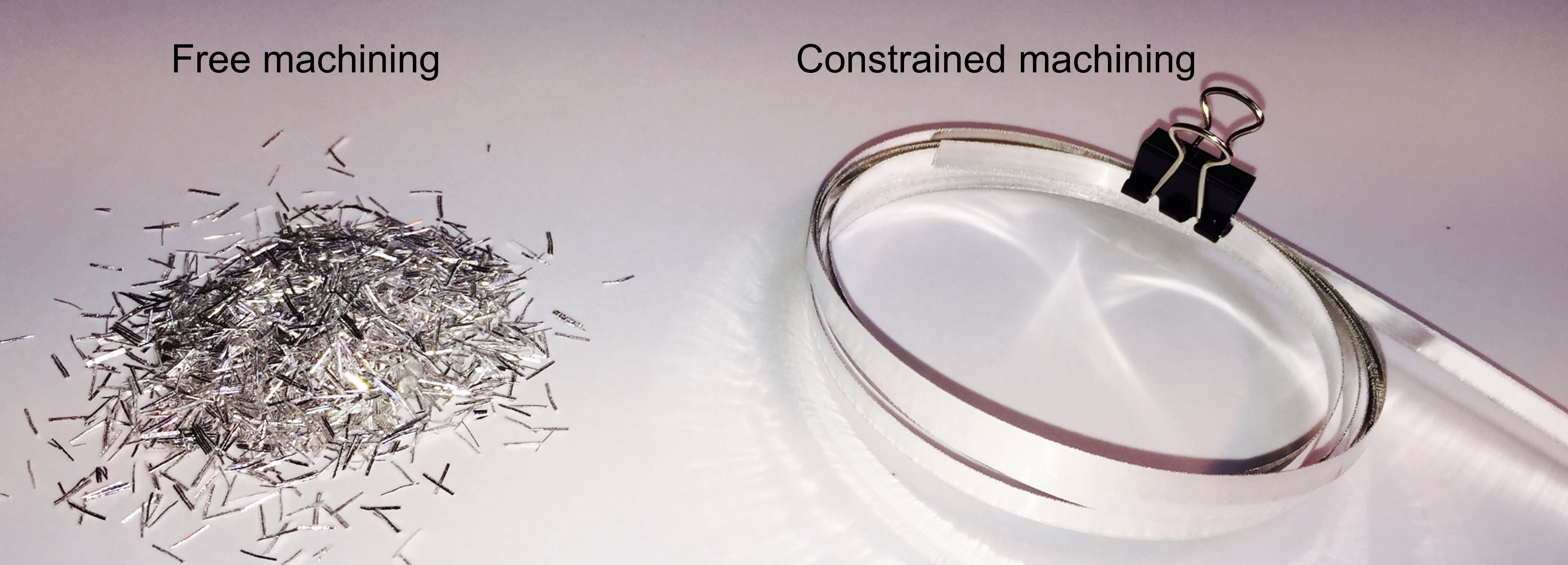}
  \caption{}
  \end{subfigure}
  
  \begin{subfigure}[b]{1\textwidth}
  \centering
  \includegraphics[width=0.8\textwidth]{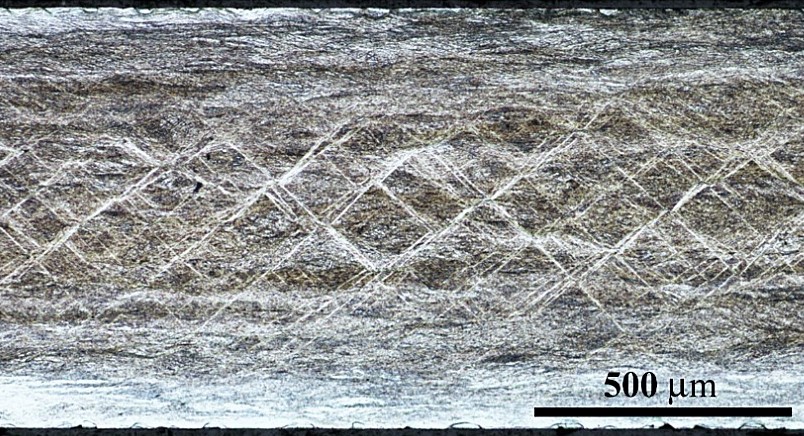}
  \caption{}
  \end{subfigure}
  
  \caption{Suppressing shear bands: (a) shows the transition in chip formation mode in machining of Mg alloy AZ31B from discrete chip formation in conventional machining (shown on left) to formation of a continuous long strip under constrained machining (right) \cite{sagapuram2016}. This transition is due to suppression of the shear banding and associated fracture via application of a geometric/physical constraint to shear band propagation. (b) Optical micrograph showing an example of how shear bands can be arrested in cold-rolling of 300M martensitic steel using compositional gradients \cite{azizi2018}. In this case, surface layers of the sample, which are characterized by a lower carbon content compared to the center, act as an effective constraint to band propagation.}
  \label{fig:control}
\end{figure}

A promising approach for suppressing shear localization is by inhibiting post-initiation shear band development, wherein much of the strain imposition and strength collapse actually occurs. This may be achieved, for example, through the application of an external kinematic constraint to arrest the flow. The practical utility of this approach has been dramatically demonstrated in machining of \emph{hcp} metals like magnesium and titanium which fail by a shear band-fracture transition at large strains. For example, Fig.~\ref{fig:control}(a) shows a side-by-side comparison of machining Mg AZ31 alloy with and without such a constraint. On the left are discrete chip particles typical of conventional machining resulting from unconstrained shear localization and subsequent fracture. This represents the most extreme case of shear banding discussed earlier in Fig.~\ref{fig:sb_initiation}, where fracture along each individual band results in the formation of discrete chip particles. On the other hand, the chip on the right in Fig.~\ref{fig:control}(a) is the result of machining under otherwise identical conditions save for the application of a second die placed directly across from the cutting tool edge. This constraint is specifically placed so as to prevent localized sliding following band initiation. Remarkably, the application of this constraint results in the discrete chips being replaced by a long continuous ribbon chip.   Metallographic examination of the ribbon also revealed a homogeneous microstructure that is devoid of shear bands or cracks   \cite{sagapuram2016}. Since the kinematics of the flow field are largely independent of microstructure (\emph{cf}. Fig.~\ref{fig:DSKVviscosity}), this purely geometric approach also applies equally well to other alloy systems (e.g., Ti and Ni-based alloys).  

It should be also noted that the constraint to shear band propagation does not always have to come from an external member. A recent study has shown how internal composition gradients can be used to control band development in rolling of martensitic steels \cite{azizi2018}. In this study, surface decarburization was used to generate $\sim 0.5$ mm thick surface layers of lower carbon composition ($\sim 0.1$\%), and consequently lower hardness/brittleness, when compared to the bulk with $\sim 0.4$\% carbon content. Figure~\ref{fig:control}(b) shows a cross-sectional view of this material after cold rolling, where shear bands which have nucleated in the center are seen to be arrested at the decarburized surface layers. This enabled large thickness reductions (equivalent strain $> 2$) to be achieved in this martensitic steel, which is otherwise known to be virtually impossible to cold roll due to banding-induced catastrophic fracture. Other attempts have also been made, such as altering  the propagation behavior of nanometer-scale shear bands in metallic glasses via introduction of ductile crystalline phases \cite{hays2000,donohue2007}. However, extension of these ideas to demonstrate macroscale control remains a technological challenge.

In complete contrast to these strategies aimed at shear band prevention, one may well envisage situations where shear localization is, in fact, beneficial. One such application, outside of the ambit of materials processing, that has received some attention in the past is in terminal ballistics, where the projectile's penetration capability is determined by the stability of plastic flow at its head. In this case, the occurrence of plastic flow localization (along two orthogonal planes of maximum shear) and subsequent discard of failed material at the head of the projectile has been shown to actually improve the penetration capacity by enabling the projectile head to maintain an acute shape---so-called \lq self-sharpening\rq mechanism \cite{magness1994}. Using simple energy considerations, it can be argued that shape changes mediated by flow localization along thin shear bands likely require less energy when compared to uniformly spread-out stable plastic flow, therefore projectiles that exhibit shear localization should be able to achieve a greater penetration depth for a given kinetic energy. By extension, flow localization may also be exploited to carry out material separation or fracture in industrial processes such as blanking and machining with reduced forces and energy \cite{latanision1983}. Although this was suggested by Zener for punching of steels \cite{zener1944a}, this possibility is only now beginning to receive significant attention \cite{viswanathan2017,healy2015}.

Another area that seems virtually unexplored, and where shear bands can be potentially put to use, is perhaps in crystallographic texture engineering for sheet metal rolling. Crystallographic texture, which affects many properties including elastic modulus, sheet formability and electromagnetic properties, is intricately linked to the material's deformation history during processing. Therefore, significant scientific effort has been profitably spent on engineering particular textures (via optimizing processing/heat treatment schedules) to produce sheet materials with specific characteristics or behavior. 
Given that the localized deformation within shear bands (uniaxial simple shear) is inherently different from that of adjoining low-strain regions, this suggests opportunities to effect new crystal orientations along these bands which can be subsequently made to recrystallize and grow into the neighboring matrix. Thus, one can possibly engineer new textures that are otherwise not possible to achieve via stable plastic flow alone. Even though the critical role of shear band deformation in the overall texture evolution has been quite extensively studied in the context of electrical steels \cite{haratani1984,humphreys2012}, their application as a general texture engineering \lq tool\rq\ has not received attention. This is perhaps because of current limitations in \emph{a priori} controlling shear band attributes such as band spacing, strain and orientation, while also ensuring that the material along the band does not catastrophically fail---a serious technical challenge indeed. However, given the urgent need for creating novel textures in many lightweight sheet products, particularly \emph{hcp} magnesium alloys \cite{hirsch2013}, the time seems opportune for focused scientific investigation.

\section{Summary and some concluding remarks}
\label{sec:discussion}

In reviewing the experimental and theoretical results of the preceding sections, one might be led to believe that the data is too scattered and diffuse to draw any general conclusions. While this is partly true, due primarily to the widespread occurrence of shear bands in different material systems, and across length scales and deformation geometries, it should be clear by now that the process of intense shear localization along a well-defined band can be delineated into distinct phases. Firstly, in the band initiation phase, a weak interface is formed  either due to a continuum-scale instability from adiabatic heating or mediated by other microstructural factors (Sec.~\ref{sec:shearBandInitiation}). Secondly, subsequent deformation occurs along this initiated interface and remains confined to a very narrow zone, analogous to a boundary layer. Finally, once straining has ceased, but the loading continued, void growth may occur inside the band leading to catastrophic fracture. This delineation is also instructive in identifying some outstanding issues pertaining to the individual phases and addressing them separately. In our opinion, despite much progress on characterizing shear banding using sophisticated experimental (e.g., \emph{in situ} methods, high-resolution microscopy) and computational techniques, several theoretical difficulties remain answered. 

To begin with, we are far from enumerating \emph{all} possible micromechanisms for shear band initiation. Even with the existing theories, primarily either thermal or mechanical in origin, the precise material/loading conditions under which these mechanisms become important remain unclear. Furthermore, these theories themselves are still fraught with basic questions. For instance, thermally activated softening mechanisms such as dislocation climb, dynamic recrystallization and phase transformations  are all diffusion-controlled, and therefore often are incompatible with the small time scales associated with band formation. On the other hand, the picture that shear band nucleation is a mechanical instability of the crystal lattice due to sudden breakdown of dislocation barriers needs obvious addressing in order to explain how dislocation motion on well-defined crystal planes within a grain can result in macroscopic bands that are apparently insensitive to the microstructure. Perhaps, this is part of a wider problem concerning how microscopic flow carriers such as dislocations collectively organize across large distances so as to produce macroscopic deformation patterns (e.g., slip bands, PLC bands). Possibly related questions often arise in the study of collective excitations in condensed matter physics, such as the occurrence of coherent defect structures during deformation of disordered metallic glasses \cite{langer2006,greer2013}. Moreover, the mechanisms responsible for dislocation locking can be wide-ranging, for example, mechanical (precipitates, solute atoms, grain boundaries), chemical or electronic in nature; these have not been systematically analyzed. On the other extreme, the theory of texture softening where localization is caused by lattice rotation into a geometrically softer orientation appears to be clear of kinetics issues, and yet this theory is generally applied only to quasi-static or low strain-rate deformation conditions and highly pre-strained metals \cite{dillamore1979}.

When it comes to the issue of post-initiation shear band development, our understanding of the effects of material properties and deformation conditions on the band front geometry and propagation velocity remains sketchy at best. For example, contradicting reports exist on whether the propagation velocity is a material-dependent constant or purely determined by external strain rate \cite{zhou1996,yadav2019}. It may be that the precise mechanism driving shear band propagation also governs the band front properties. In a similar vein, other than the observation that shear bands cut across microstructural features such as grain or twin boundaries with little constraint, their interactions with potentially effective obstacles such as secondary ductile phases are not yet understood. This, in turn, means that we are equipped with limited background for engineering tools to arrest or control shear band propagation.

It is perhaps in the area of high strain/strain-rate plastic flow associated with shear band growth that most progress has been made, owing in part to direct \emph{in situ} measurements of displacement/strain fields around an evolving shear band, combined with high-resolution microstructural analysis. However, here too several outstanding problems are noteworthy. Thermal diffusion away from the band is all too often attributed as the primary factor governing band thickness. Careful experimental analyses \cite{sagapuram2018} on the other hand show that  viscous effects are likely far more important, even though our understanding of the factors influencing the local shear band viscosity and constitutive behavior is still nascent. As regards to the microscopic mechanism, the notion that the viscous behavior at shear bands arises due to phonon damping of dislocations  warrants quantitative investigation and further experimental confirmation across different materials and strain rates.

In spite of the rather large number of experimental investigations of shear bands, very few remain dedicated to  critical  examination of fundamental mechanisms underlying single band formation. This is likely because most observations of shear bands have been made incidental to some other main objective. Similarly, one can cite many examples where the occurrence of shear banding is merely noted with little to no attempt made to investigate the dynamics of band formation or understand the effects of different experimental factors. For example, and as noted earlier, there has been little systematic study concerning the effects of material composition/microstructure or imposed deformation rate on shear band characteristics such as propagation kinematics, band width, strain/temperature profiles and microstructure. It should be obvious that in view of the current state of theoretical deficiencies, there is an urgent need for such investigations.

In our view, basic decoupling of the shear banding process into distinct stages of initiation, development and (potentially) fracture provides a logical experimental framework in which to critically evaluate alternative mechanisms/theories, allowing a reasonably complete picture of this complex problem to be built. In this regard, and in the context of direct \emph{in situ} observational studies, high-rate deformation configurations such as punching and cutting which effect nucleation of a single, well-defined shear band at an \emph{a priori} known location should prove most useful. The simultaneous initiation of shear bands at multiple nucleation sites and their mutual interactions, such as in explosively loaded thick-walled cylinders \cite{xue2002,lovinger2015,navarro2018}, is an even more involved problem that is largely outside the range of the present review.

Caution must be exercised when undertaking purely microstructural studies of shear bands, as is generally common within the materials science community. However sophisticated, such investigations by themselves cannot reveal much about band nucleation mechanisms. This is because any \lq fingerprint\rq\ of the processes preceding and responsible for band nucleation are usually obscured by the intense plastic flow and attendant microstructural changes occurring post-instability. On the other hand, one can draw great benefit from detailed microstructural studies as a function of the local shear band strain/strain rate/temperature histories, as they will serve to reveal how metals generally behave under extreme thermomechanical conditions. Similarly, fractography investigations taken up in conjunction with the \emph{in situ} dynamics studies can help relate observed fracture morphologies to shear band deformation history, and should prove to be most valuable.

One area of  study that hold much promise is the field of low melting point (fusible) alloys ($T_m \leq$  200 $^\circ$C). It has recently been demonstrated shear bands can be initiated in these systems at strain rates that are multiple orders of magnitude lower compared to typical rates encountered in conventional alloys. This means that all the essential features of shear banding, including band initiation and development, growth dynamics and transition to fracture can be captured \emph{in situ} without loss of any temporal or spatial resolution. We have recently utilized one such alloy (Wood's metal) to characterize shear band propagation velocities and to further resolve the spatiotemporal evolution of plastic flow around a growing band \cite{yadav2019,yadav2020}. As a natural extension, this allows combining kinematic displacement/velocity data with simultaneous temperature measurements, as discussed earlier.  Replicating shear banding in polycrystalline metals at quite low strain rates using low melting point alloys (correspondingly, high homologous temperatures) parallels the concept of time-temperature superposition which is well-known in polymers, but has not been investigated much in metals. Thus, low melting point alloys present an interesting possibility to experimentally \lq simulate\rq\ the high strain-rate behavior of structural metals, including shear banding, at quite small deformation speeds. They also provide a convenient framework wherein the effects of deliberate alloying and microstructural alterations on band dynamics can be studied systematically.

Lastly, it should be pointed out that machine stiffness and related elastic energy effects have been largely sidelined over the past several years, even though the original investigations of Basinski and others \cite{basinski1957, chin1964a,lemaire1972} have clearly demonstrated their importance. Shear band studies with \lq soft\rq\ machines possessing controlled and variable stiffness  could be useful in addressing questions pertaining to band nucleation. Similarly, stress-controlled, as opposed to conventional displacement-controlled, experiments have produced many interesting observations in the field of L\"{u}ders/PLC-like plastic instabilities \cite{abbadi2002}, and may be another avenue worth exploring in the context of shear bands.

It is hoped that these guidelines will help advance the state-of-the-art in this important area of large strain deformation of metals, and equip the mechanics community in better understanding the behavior of next-generation high strength material systems.


\appendix
\section{Appendix}
\label{sec:appendix}
The following excerpt on the 1944 Zener-Hollomon paper is taken from Zener's interview by Lillian Hoddeson in 1981 at the American Institute of Physics \cite{zener_interview}:

\begin{displayquote}
At this time, a lot of laboratories were working on how the strength of steel depended on rate of deformation. We recognized that you can eliminate most of this work by varying at the same time that you vary speed, you vary
temperature, since every physicist knows that things speed up as you raise the temperature and by a very simple law---you only don't know the heat of activation of what's going on. So this (1944) paper and some of these other papers were related to demonstrating how you can obtain much more for a given amount of work if you work at different temperatures $\ldots$

This (work) I wish to emphasize particularly since something very important in general came out of this. A lot of work (that) was being done on the effect of strain rate  missed the main point: The effect of strain rate does in fact influence the stress level, perhaps 10 or 20\%,  but even more importantly, the difference in going from low speed isothermal to adiabatic deformation can make a tremendous qualitative difference in behavior, and that's what this (paper) is about. These papers are so long I don’t know how we wrote so much. I mean, we weren't being paid by the word. As you keep on straining the mixture at a given strain rate, the stress required to produce further strain goes up, so to speak, (the material) hardens. This is what happens under isothermal conditions, but again all through my life I have been working with adiabatic and isothermal, and the differences between the two. Now with adiabatic deformation you have the strain hardening, true, but you also have the temperature rise. Temperature per se weakens the material, so you are also having this weakening effect of temperature induced by deformation. So, if you plot a isothermal stress-strain curve, it keeps on going up. But, adiabatically you plot it, it would be the same at first but the deviation will get larger and larger and you will rich a maximum $\ldots$ Once you reach this maximum then this (strain) goes up suddenly and if it's (material) sheared, you will shear it off just along one plane---or as Seitz and Bardeen would say, dislocations are involved along one plane.

And then, we did experiments to demonstrate that and these experiments I am very proud of. These experiments where you have a plate supported by a die, and then if you just push slowly, it (plate) deforms and then tears. But, if you drop a weight on it, not so it goes through, but so it goes when you take a section (of the plate, you see a thin white layer).  The white means that it has become so hot, all further deformation is concentrated. So the essential thing is in going to higher speeds, but once the speed is high enough so it is adiabatic deformation, that's the major effect, because then it changes the whole pattern of deformation. So and then we show that rather than having a die, if you simply have a plate and fire a projectile, that the same thing happens.

\end{displayquote}



\bibliography{bibfile}
\bibliographystyle{unsrt}
\end{document}